\documentclass{WileyMSP-template}
\usepackage{xcolor}
\usepackage{times}
\usepackage{cite}
\usepackage{graphicx}
\usepackage{amsmath}
\usepackage{amssymb}
\usepackage{siunitx}
\RequirePackage{etoolbox}
\usepackage{hyperref}
\usepackage[nameinlink,noabbrev,capitalize]{cleveref}
\usepackage{xr-hyper}
\DeclareFontEncoding{LS1}{}{}
\DeclareFontSubstitution{LS1}{stix2}{m}{n}
\DeclareMathAlphabet{\mathscr}{LS1}{stix2scr}{m}{n}
\begin{document}

\pagestyle{fancy}

\title{Symmetry Breaking in Chemical Systems: Engineering Complexity through
	Self-Organization and Marangoni Flows}

\maketitle


\author{Sangram Gore,} 
\author{Binaya Paudyal,} 
\author{Duarte Rocha,}
\author{Mohamed Ali,}
\author{Nader Masmoudi,}
\author{Albert Bae,}
\author{Christian Diddens,}
\author{Detlef Lohse,}
\author{Oliver Steinbock,}
\author{Azam Gholami*}


\dedication{}

\begin{affiliations}
Dr. S. Gore, B. Paudyal, Dr. M. Ali, Prof. N. Masmoudi, Prof. A. Gholami\\
Science Division, New York University Abu Dhabi, Abu Dhabi, UAE\\
Email Address: azam.gholami@nyu.edu\\

D. Rocha, Dr. C. Diddens, Prof. D. Lohse \\ 
Physics of Fluids Department, Max-Planck Center Twente for Complex Fluid Dynamics and
J. M. Burgers Centre for Fluid Dynamics, University of Twente, Enschede, The Netherlands\\
Prof. D. Lohse \\ 
Max-Planck Institute for Dynamics and Self-Organization, Am Faßberg 17, 37077 Göttingen, Germany\\
Prof. N. Masmoudi\\
Courant Institute of Mathematical Sciences, New York University, New York, USA

Prof. A. Bae\\
Lewis $\&$ Clark College, Portland, Oregon, USA

Prof. O. Steinbock\\
Department of Chemistry and Biochemistry, Florida State University, Tallahassee, Florida, USA

\end{affiliations}



\begin{abstract}

Far from equilibrium, chemical and biological systems  can form complex patterns and waves through reaction-diffusion coupling. Fluid motion often interferes with these self-organized concentration patterns. In this study, we investigate the influence of Marangoni-driven flows inside a thin layer of fluid ascending the outer surfaces of hydrophilic obstacles on the spatio-temporal dynamics of chemical waves in the modified Belousov-Zhabotinsky reaction. Our observations reveal that circular waves originate nearly simultaneously at the obstacles and propagate outward. In a covered setup, where evaporation is minimal, the wavefronts maintain their circular shape. However, in an uncovered setup with significant evaporation and resulting Marangoni flows, the interplay between surface tension-driven Marangoni flows and gravity destabilizes the wavefronts, creating distinctive flower-like patterns around the obstacles. Our analysis shows that here solutal Marangoni forces are more relevant than thermal ones. Our experiments further show that the number of petals formed increases linearly with the obstacle's diameter, though a minimum diameter is required for these instabilities to appear. These findings demonstrate the potential to 'engineer' specific wave patterns, offering a method to control and direct reaction dynamics. This capability is especially important for developing microfluidic devices requiring precise control over chemical wave propagation.

\end{abstract}


\setlength{\parindent}{15pt}
\setlength{\parskip}{0pt}

\section{Introduction}
Symmetry breaking is a fundamental concept in physics, biology and chemistry where small perturbations can lead to the emergence of complex
structures and patterns from initially uniform states~\cite{hoyle2006pattern,meinhardt1992pattern}. In fluid dynamics, this phenomenon is exemplified by the transition from simple,
laminar flows to intricate turbulent structures, driven by instabilities
that disrupt the symmetry of the system. In biology, Alan Turing's reaction-diffusion theory proposes that the initial symmetry in embryos can be broken by the interaction of two diffusible molecules, whose dynamic interplay leads to the formation of complex patterns~\cite{turing1990chemical}. Similarly, in chemical systems
like the Belousov-Zhabotinsky (BZ) reaction, symmetry breaking can give
rise to a rich spectrum of spatio-temporal structures, such as rotating
spirals and target waves as well as Turing patterns~\cite{zaikin1970concentration,epstein1998introduction,field1974oscillations,tyson2013belousov,steinbock1993control,winfree1974rotating,steinbock1995navigating,ginn2004quantized,ginn2004microfluidic,steinbock2013wave,suzuki2000diffusive,mikhailov2006control,agladze1994rotating,guo2010spontaneous,zykov2018spiral,zykov1987simulation,duzs2019turing,yashin2006pattern,taylor2015insights,taylor1999scroll}. These patterns, resulting from the interplay between nonlinear reaction kinetics and diffusion, offer profound insights into the mechanisms of
self-organization in both natural and synthetic systems~\cite{hoyle2006pattern,halatek2018rethinking,epstein2006introduction}.

The interaction between chemical reactions and convective fluid motion, particularly through mechanisms such as Marangoni-driven flows, profoundly influences the formation and breakdown of complex spatio-temporal patterns~\cite{de2020chemo, budroni2017dissipative,bigaj2024thermal,bigaj2023marangoni,lohse2022fundamental}.  In pioneering studies on the BZ reaction~\cite{showalter1980pattern,agladze1984chaos,muller1988chemical,miike1988oscillatory2,miike1988oscillatory}, it was observed that in confined environments, chemical instabilities manifest themselves as propagating waves. In contrast, in systems with a free surface, the presence of Marangoni flows lead instead to the formation of cellular structures~\cite{kai1994hydrochemical,kai1995curious,inomoto1995depth,epstein2012chemical,vanag2009cross, rossi2012segmented, rossi2009chemical,miike2010flow,matthiessen1995global,matthiessen1996influence,budroni2009bifurcations,ginn2004microfluidic}. These flows are initiated by gradients in surface tension~\cite{yoshikawa1993generation,matthiessen1996influence}, which might arise from either temperature variations  or differences in solute concentrations  across the fluid's surface, causing the fluid to flow from regions of lower to higher surface tension.  For instance, in the ferroin-catalyzed BZ reaction, the Fe$^{2+}$ complex has a higher surface activity than the Fe$^{3+}$ complex which is attributed to the differing hydrophilic properties of the catalyst's oxidized (more hydrophilic) and reduced forms. This difference leads to a decrease in surface tension of approximately 2 mN/m when the system transitions from the reduced to the oxidized state~\cite{yoshikawa1993generation}. In an uncovered setup, where evaporation is significant, the interplay of chemical waves, buoyancy-driven flows due to density gradients, and thermal and solutal Marangoni flows due to surface tension gradients collectively influences the resulting chemo-hydrodynamic patterns~\cite{agladze1984chaos,plesser1992interaction,wilke1995interaction,rossi2009chemical,rossi2012segmented,rongy2006steady,miike1988oscillatory,lohse2022fundamental}.  

Evaporation-induced Marangoni flows can be intensified by introducing hydrophilic vertical or inclined obstacles within the reactive medium. The thin film rising along the outer surfaces of these obstacles evaporates more quickly than the bulk fluid, creating thermal and/or solutal surface tension gradients by evaporative cooling and/or different volatilities of the multicomponent liquid. The delicate balance between upward-driving surface tension forces and gravity-induced downward motion leads to undulations  at the fluid interface, potentially developing into finger-like patterns if the fluid becomes sufficiently unstable~\cite{oron1997long,bonn2009wetting,craster2009dynamics,bertozzi1998contact,bertozzi1999undercompressive,munch1999rarefaction,smolka2017dynamics}. The obstacle's geometry also affects the dynamics at the droplet-gas of thin films \cite{smolka2017dynamics, de2021marangoni}. A cylindrical obstacle introduces unique conditions, such as azimuthal curvature and periodic boundary conditions, absent in flat obstacles~\cite{smolka2017dynamics}. This periodicity ensures that an integer number of fingers can form along a cylinder, a condition that is not required for a flat obstacle, and the smallest number of fingers that can form along the contact line of a cylinder is one~\cite{smolka2017dynamics}.

Building on these groundworks, our study introduces a novel approach to explore the intricate interaction between evaporation-driven Marangoni flows and chemical waves in the modified BZ reaction around hydrophilic obstacles. In a covered setup -- where evaporation and therefore the evaporation-induced surface tension differences are minimal -- these obstacles act as focal points, emitting synchronous circular waves that retain their shape over multiple cycles. By contrast, in an uncovered setup with pronounced evaporation, thermal and solutal Marangoni flows trigger a hydrodynamic instability, yielding characteristic flower-like wave patterns. The number of ‘petals’ scales with obstacle diameter, and a critical diameter threshold is required for the instability to emerge. \textcolor{black}{ We also show that in our system, solutal Marangoni forces are more relevant than thermal ones. To elucidate the driving mechanism, we perform numerical simulations of a simplified binary fluid (e.g., water–ethanol) subject to solutal Marangoni instabilities, due to selective evaporation of one component (e.g.\ ethanol as compared to water). These simulations qualitatively reproduce the observed symmetry-breaking patterns and lay the groundwork for future quantitative modeling. }
\section{Results}
\subsection{Experimental Results}
\begin{figure}[t!]
	\begin{center}
		\includegraphics[width=\columnwidth]{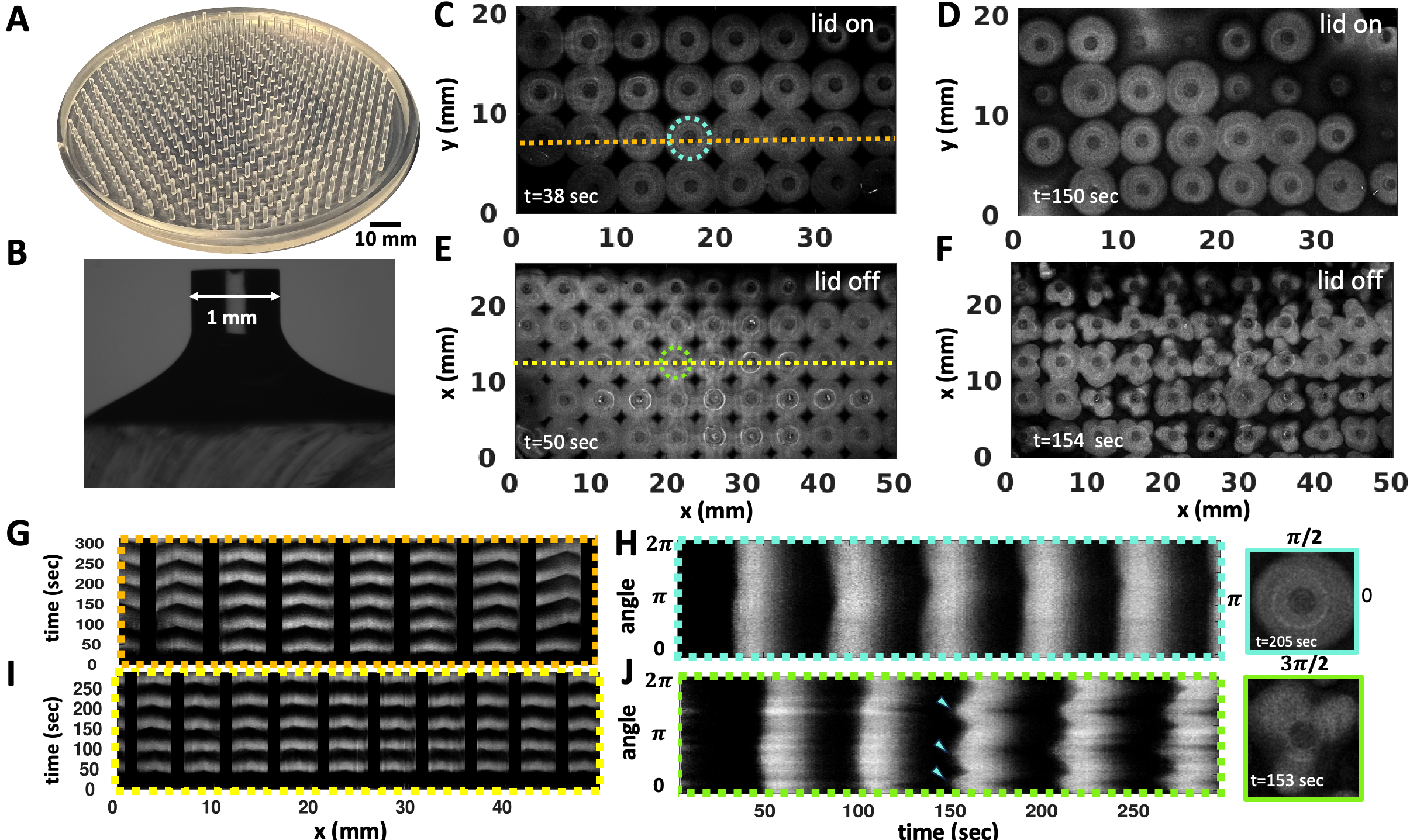}
		\includegraphics[width=0.98\columnwidth]{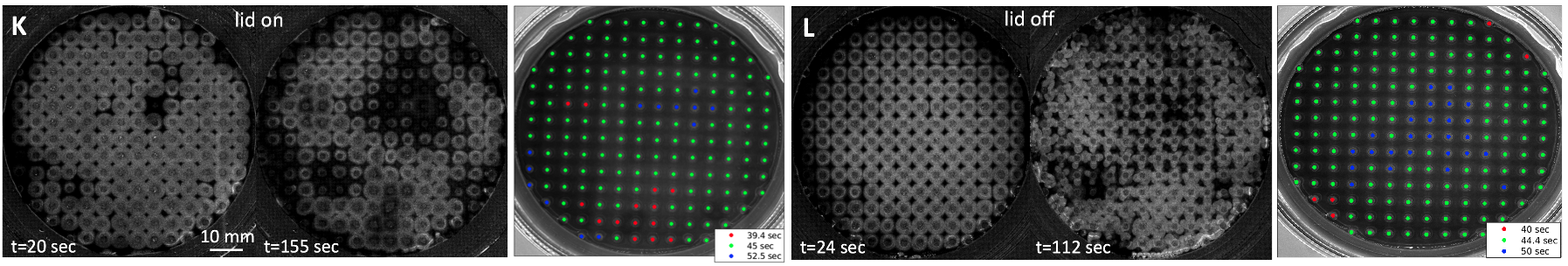}
		\caption{\textbf{Wave dynamics in a covered versus an uncovered setup}.  (\textbf{A}) A standard experimental setup featuring a periodic arrangement of PDMS-based cylindrical obstacles designed to fit within a 10 cm Petri dish. (\textbf{B}) A side view depicting the fluid profile around an exemplary hydrophilic pillar. (\textbf{C}-\textbf{D}) Top views of concentric chemical waves in a covered setup, captured at two distinct times, show the waves initiating almost simultaneously at the pillars and spreading outward. The waves retain their circular shapes over time. (\textbf{E}-\textbf{F}) Top views of chemical waves captured at two distinct times in an uncovered setup: initially, concentric waves form around the obstacles (E), and later, the wavefronts fracture to create flower-like patterns (F). (\textbf{G}) A space-time plot along the orange line indicated in (C) demonstrates that the pillars serve as the wave centers. Black bars show the position of the obstacles. (\textbf{H}) A space-time plot traced along the cyan-colored circle shown in (C) illustrates that the circular waves retain their shape through multiple wave cycles. (\textbf{I}) The space-time plot along the yellow  line shown in (E). (\textbf{J}) The space-time diagram along the green circle indicated in (E) illustrates that after two cycles of circular waves, the wavefronts break. In this experiment, where pillars are  1 mm in diameter, 3 mm tall, and spaced 5 mm apart, most pillars develop three petals.  (\textbf{K}-\textbf{L}) Images displaying the entire Petri dish from two separate experiments: one covered and the other uncovered. The color-coded wave period around each pillar is also shown in seconds. The timestamps reflect the elapsed time since the start of recording, which began shortly after the chemical solution was introduced into the setup. These images are typical of at least 50 experiments. }
		\label{fig:PetriCloseOpen}
	\end{center}
\end{figure}

Our quasi 2D setup consists of a periodic array of millimeter-sized PDMS-based cylindrical obstacles (see \textbf{Figure~\ref{fig:PetriCloseOpen}A}) that influence the spatio-temporal dynamics of chemical waves in the modified BZ reaction (see Method Section). This version of the BZ reaction, known as CHD-BZ, employs 1,4-cyclohexanedione as a bubble-free organic substrate and ferroin as a high-absorbance redox indicator~\cite{farage1982uncatalysed,szalai2002mechanistic,bansagi2006nucleation,hamik2001anomalous,manz2003meandering,hamik2003excitation,bansagi2008three,bansagi2007negative}. The addition of the ferroin redox indicator to the reaction mixture renders the oscillations visible, with periodic color changes between blue and red. The PDMS has undergone plasma treatment to make it hydrophilic, which allows the fluid to climb up around these obstacles, as illustrated in Figure~\ref{fig:PetriCloseOpen}B. We observe circular oxidation waves in blue color (high value of ferriin concentration Fe$^{3+}$) that initiate almost synchronously at the pillars and propagate inside the reduced state in red color (high concentration of ferroin Fe$^{2+}$) at a speed of approximately 7 mm/min and recurring every 45 sec (Figure~\ref{fig:PetriCloseOpen}C-L). When these waves collide, they mutually annihilate each other. The time evolution of these waves is strongly influenced by whether the setup is covered or uncovered (as illustrated in Figure~\ref{fig:PetriOpen_1And1_5}B), which impacts the evaporation rate of the liquid  (Figure~\ref{fig:PetriOpen_1And1_5}C). Below, we discuss the results for both setups.
\subsubsection{Covered setup: Circular waves initiate synchronously at the pillars and maintain their circular shape}
In a covered setup, chemical waves begin nearly simultaneously on the pillars, with a wave period of about 45 sec, and maintain their circular shape across multiple wave cycles (Figure~\ref{fig:PetriCloseOpen}C-D and Video 1).  A space-time plot along an arbitrary line crossing the obstacles illustrates that they serve as wave centers (Figure~\ref{fig:PetriCloseOpen}G). Additionally, to visualize the wave dynamics over several cycles, we compiled the light intensity data along a circle surrounding an arbitrary pillar, confirming that the wavefronts consistently retain their circular form (Figure~\ref{fig:PetriCloseOpen}H). Figure~\ref{fig:PetriCloseOpen}K and Video 2 present a separate experiment where the entire Petri dish is imaged, demonstrating similar wave dynamics.  Fourier analysis of the waves emanating from the obstacles indicates that the majority have a periodicity of 45 sec (see Figure~\ref{fig:frequency}A-D).
\subsubsection{Uncovered setup: Circular waves begin synchronously at the pillars and eventually evolve into flower-like patterns}
\begin{figure}[t!]
	\begin{center}
		\includegraphics[width=0.96\columnwidth]{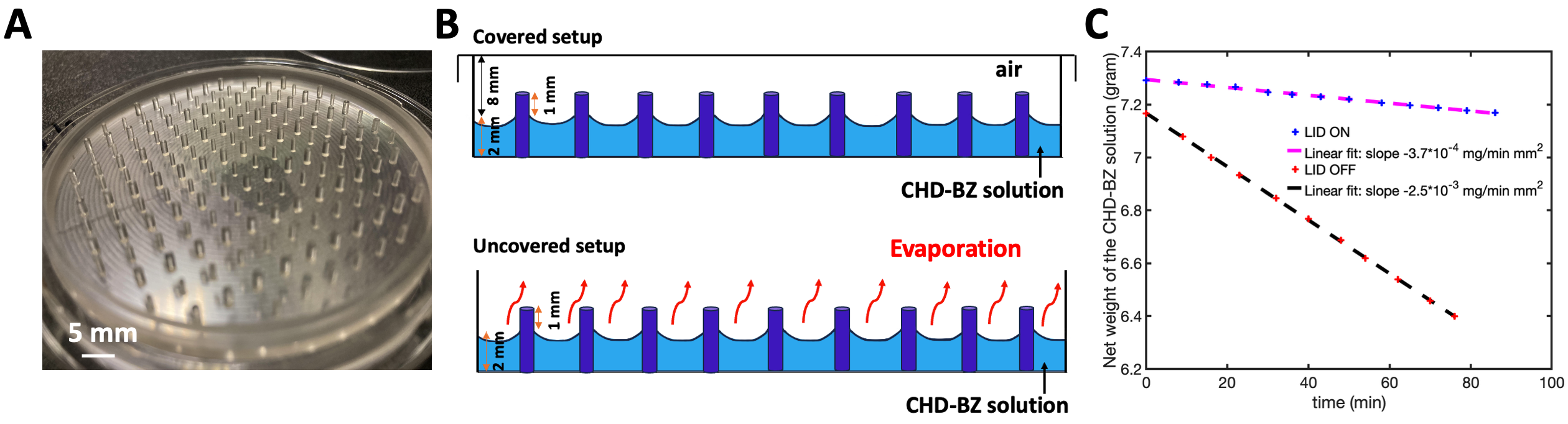}
		\includegraphics[width=\columnwidth]{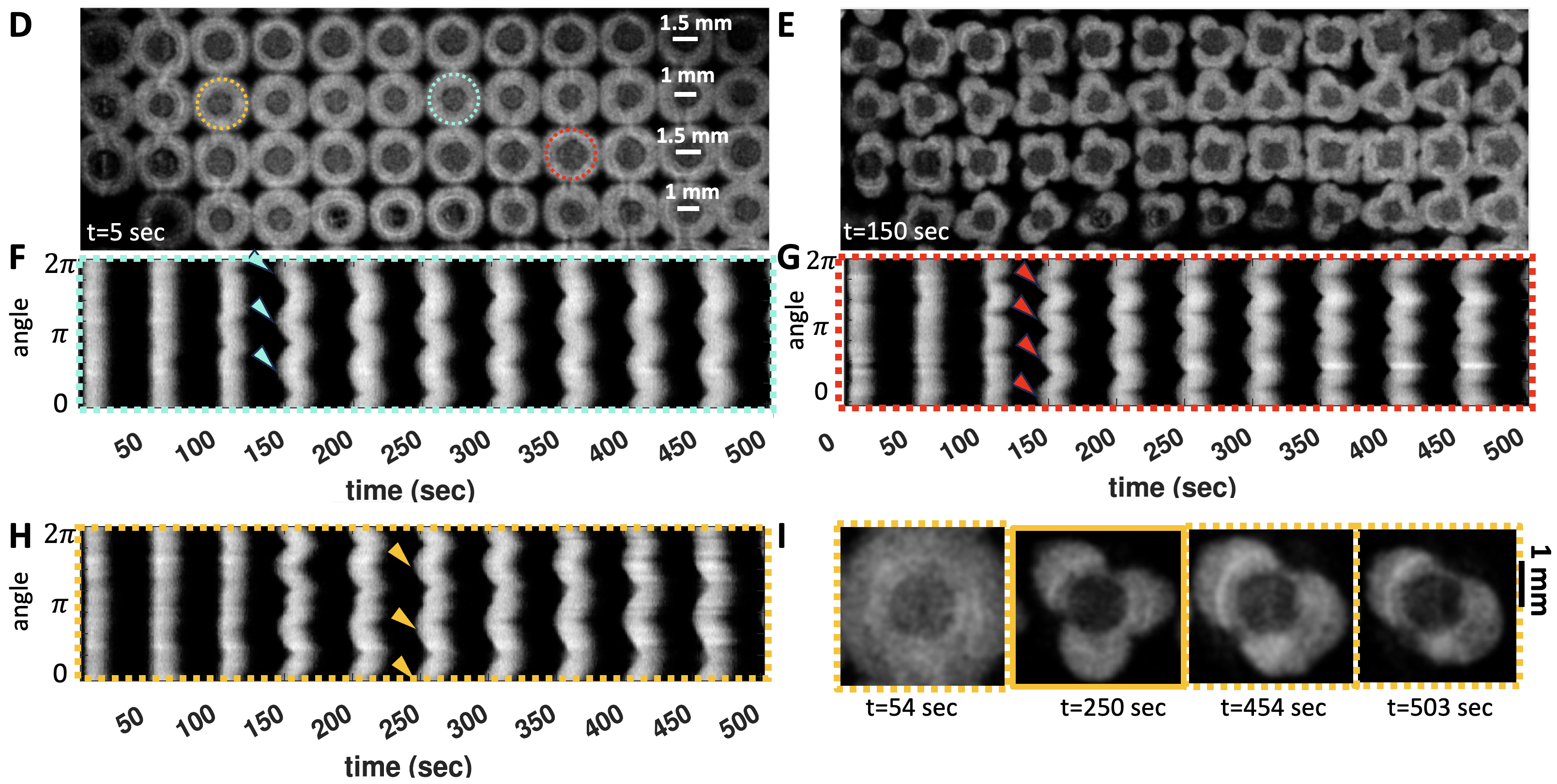}
		\caption{\textbf{Number of petals depends on the pillar diameter and can vary over time.} (\textbf{A}) A PDMS mold with alternating pillar diameters of 1 mm and 1.5 mm is shown. (\textbf{B}) A sketch illustrates the covered versus uncovered experimental setup. (\textbf{C}) In the uncovered setup, the evaporation rate is approximately 2.5$\times$10$^{-3}$ mg/(min mm$^2$), which is about seven times higher than in the covered setup. (\textbf{D}-\textbf{E}) Top views of chemical waves in an uncovered setup captured before and after the onset of instability in an experiment where pillar diameters alternate between rows, with 1.5 mm pillars in the top row and 1 mm pillars in the second row. Pillars with a diameter of 1.5 mm generally develop four petals, whereas those with a diameter of 1 mm typically produce three petals. (\textbf{F}-\textbf{H}) Space-time plots along the highlighted circles in panel (D) illustrate the progression of the instability. (\textbf{I}) An example of a pillar that initially developed three petals, but over time, two of the petals merged. }
		\label{fig:PetriOpen_1And1_5}
	\end{center}
\end{figure}
In an uncovered setup, where evaporation is more significant (Figure~\ref{fig:PetriOpen_1And1_5}C), the obstacles continue to act as centers that emit synchronous concentric waves (Figure~\ref{fig:PetriCloseOpen}E, I and Video 3). However, after approximately two cycles of these circular waves, an instability in the wavefront becomes apparent: the circular wavefronts fragment, leading to the formation of striking flower-like patterns (see Figure~\ref{fig:PetriCloseOpen}F, J). Figure~\ref{fig:PetriCloseOpen}L and Video 4 present a separate experiment where the entire Petri dish is imaged, demonstrating similar wavefront instabilities. Consistent with the covered setup, Fourier analysis shows that the waves originating from the obstacles typically have a periodicity of about 45 seconds (see Figure~\ref{fig:frequency}E-H).

\begin{figure}[t!]
	\begin{center}
		\includegraphics[width=0.9\columnwidth]{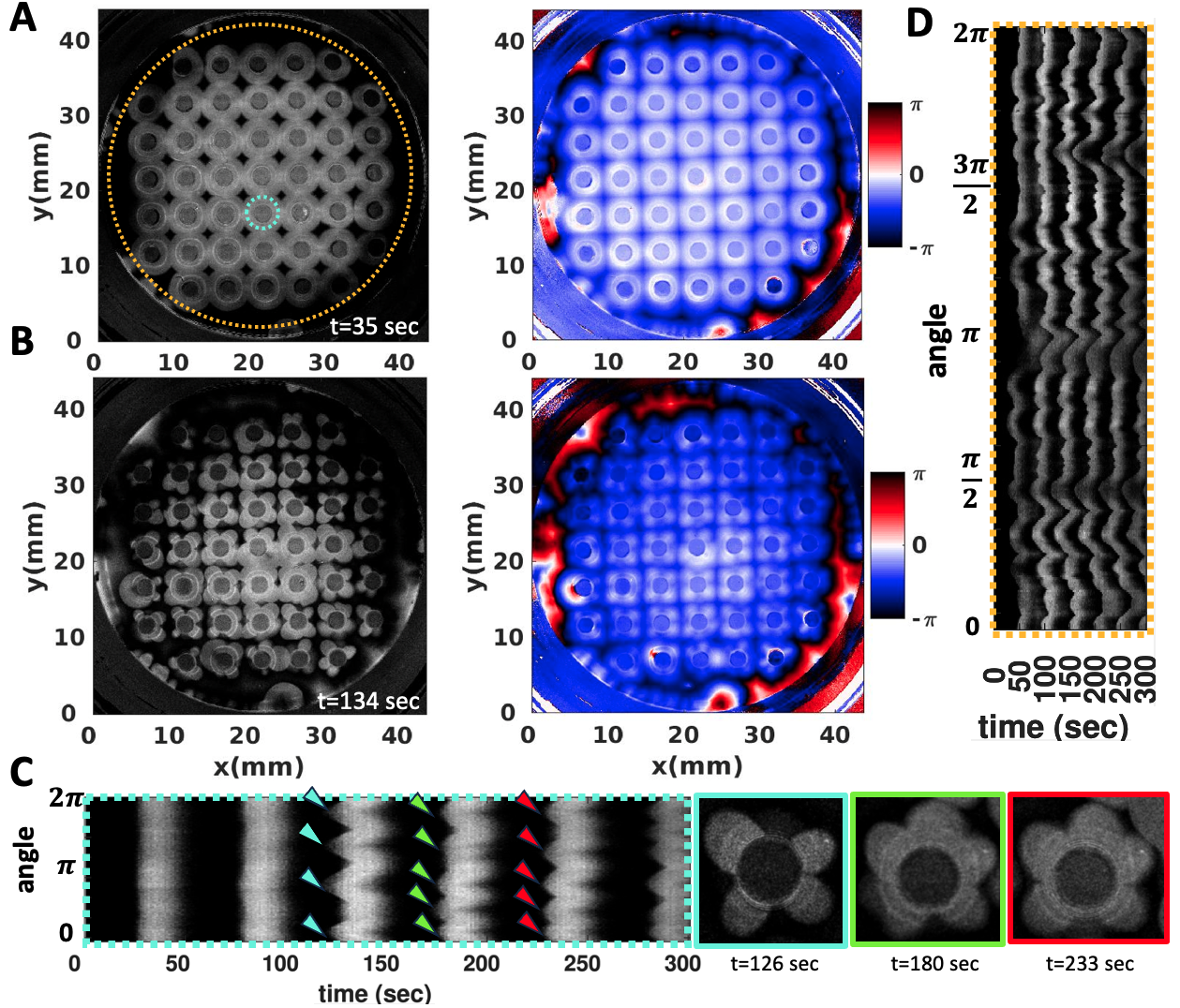}
		\caption{\textbf{Wavefront instability at the boundary in an uncovered setup.} (\textbf{A}-\textbf{B}) Pillars with a diameter of 1.85 mm typically form four petals. The color-coded phase map, generated using Hilbert transform, indicates that the pillars simultaneously act as wave centers. (\textbf{C}) A space-time plot along the cyan-highlighted circle in panel (A) shows the development of the instability, with a noticeable evolution in the number of petals over time. The wave period is 50 sec. On the right, these snapshots at three different times (\SI{126}{\second}, \SI{180}{\second} and \SI{233}{\second}) are shown. (\textbf{D}) A space-time plot along the orange-highlighted circle demonstrates the wavefront instability that also manifests at the PDMS wall on the periphery. This experiment is conducted in a small Petri dish.}
		\label{fig:PetriOpen_SmallDish}
	\end{center}
\end{figure}

\begin{figure}[t!]
	\begin{center}
		 \includegraphics[width=0.8\columnwidth]{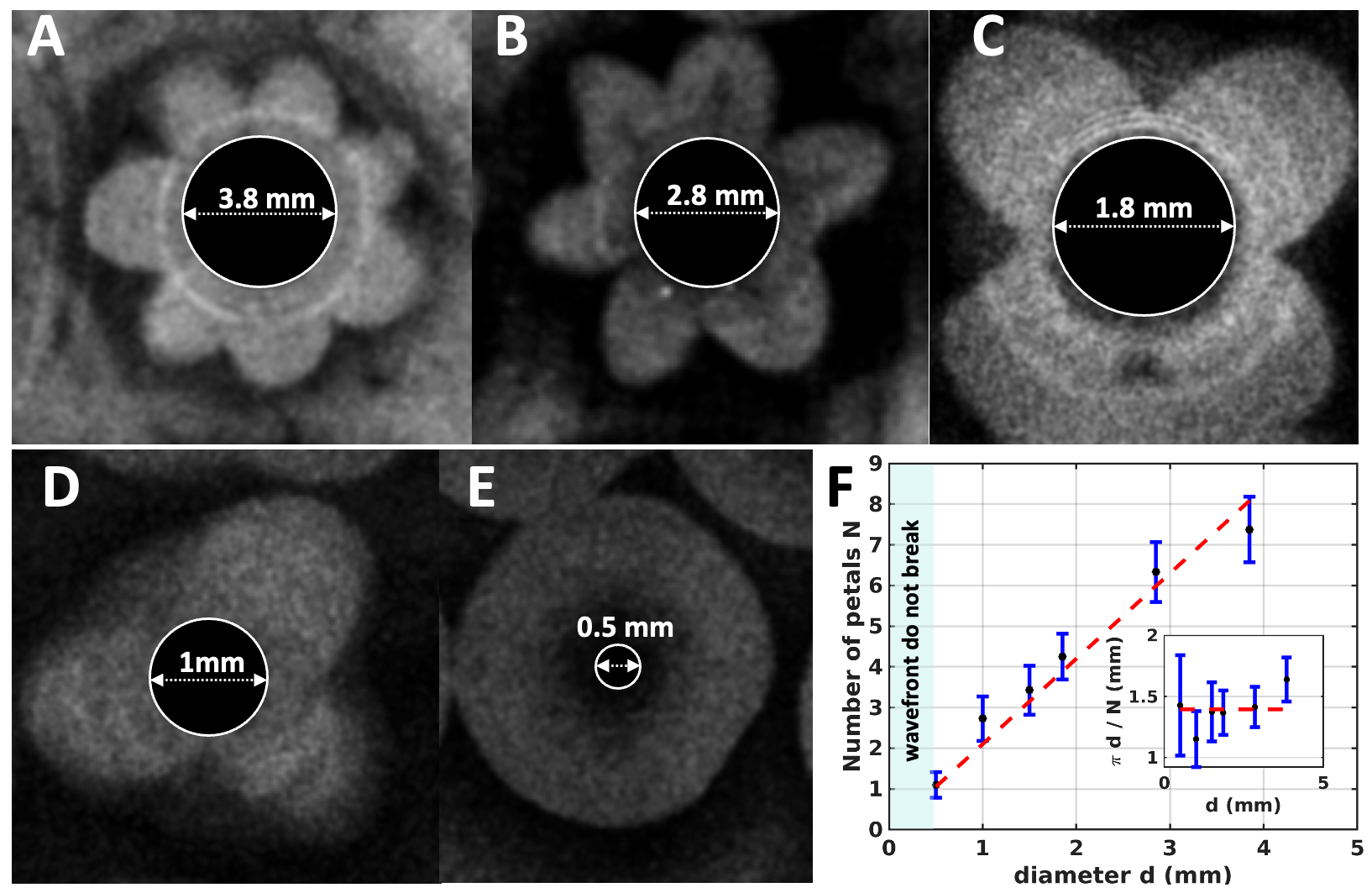}
		\caption{\textbf{Effect of the pillar diameter on the number of petals.} (\textbf{A-D}) As the pillar diameter decreases, so does the number of petals formed. For instance, pillars with a diameter of 3.8 mm typically develop around 7 petals, while those with a diameter of 2.8 mm produce about 6 petals. This number drops to 4 petals for pillars with a diameter of 1.8 mm, and further reduces to 3 petals for those with a diameter of 1 mm. (\textbf{E}) Pillars with a diameter of 0.5 mm exhibit no wavefront instability, which verifies that there is a minimum pillar perimeter required for such instability to develop. (\textbf{F}) The number $N$ of petals increases linearly with the diameter $d$ of the pillar. Below a critical diameter, wavefronts do not break and remain circular. \color{black}{The inset shows the average lateral extension $\pi d/ N$ of the petals}.}
		\label{fig:VaryingDiameter}
	\end{center}
\end{figure}

Our experimental findings indicate that the number of petals is influenced by the diameter of the pillars. Typically, pillars with a diameter of 1 mm form three petals (Figure~\ref{fig:PetriCloseOpen}F), while those with a diameter of 1.5 mm often produce four petals (Figure~\ref{fig:PetriOpen_1And1_5} and~\ref{fig:PetriOpen_SmallDish}), and pillars measuring 3.8 mm in diameter generate six to seven petals (Figure~\ref{fig:VaryingDiameter} and Video 5). Figure~\ref{fig:PetriOpen_1And1_5} presents an illustrative experiment where the pillar diameters alternate between 1 mm and 1.5 mm across neighboring rows, with the top row featuring pillars of 1.5 mm in diameter. In this setup, the 1 mm pillars typically form patterns with three petals, whereas the 1.5 mm pillars display flower patterns consisting of four petals (Figure~\ref{fig:PetriOpen_1And1_5}E-I and Video 6).

Interestingly, for a small subset of the pillars, we observed that the number of petals can vary over time, as illustrated in Figure~\ref{fig:PetriOpen_1And1_5}I. In a related observation, Figure~\ref{fig:PetriOpen_SmallDish} depicts an experiment where, after two cycles of circular wavefronts, the wavefronts around the pillars break. Initially, four petals form, but as time progresses, a fifth petal develops (Figure~\ref{fig:PetriOpen_SmallDish}C). The synchronization between the waves initiated at different pillars is demonstrated through a phase map calculated using the Hilbert transform (see Method Section), as shown in Figure~\ref{fig:PetriOpen_SmallDish}A-B. In this experiment, where we have captured images of the entire small Petri dish, the waves initiated at the PDMS boundary also exhibit wavefront instability, similar to those observed at the pillars (see Figure~\ref{fig:PetriOpen_SmallDish}D and Video 7).

\begin{figure}[t!]
	\begin{center}
		\includegraphics[width=\columnwidth]{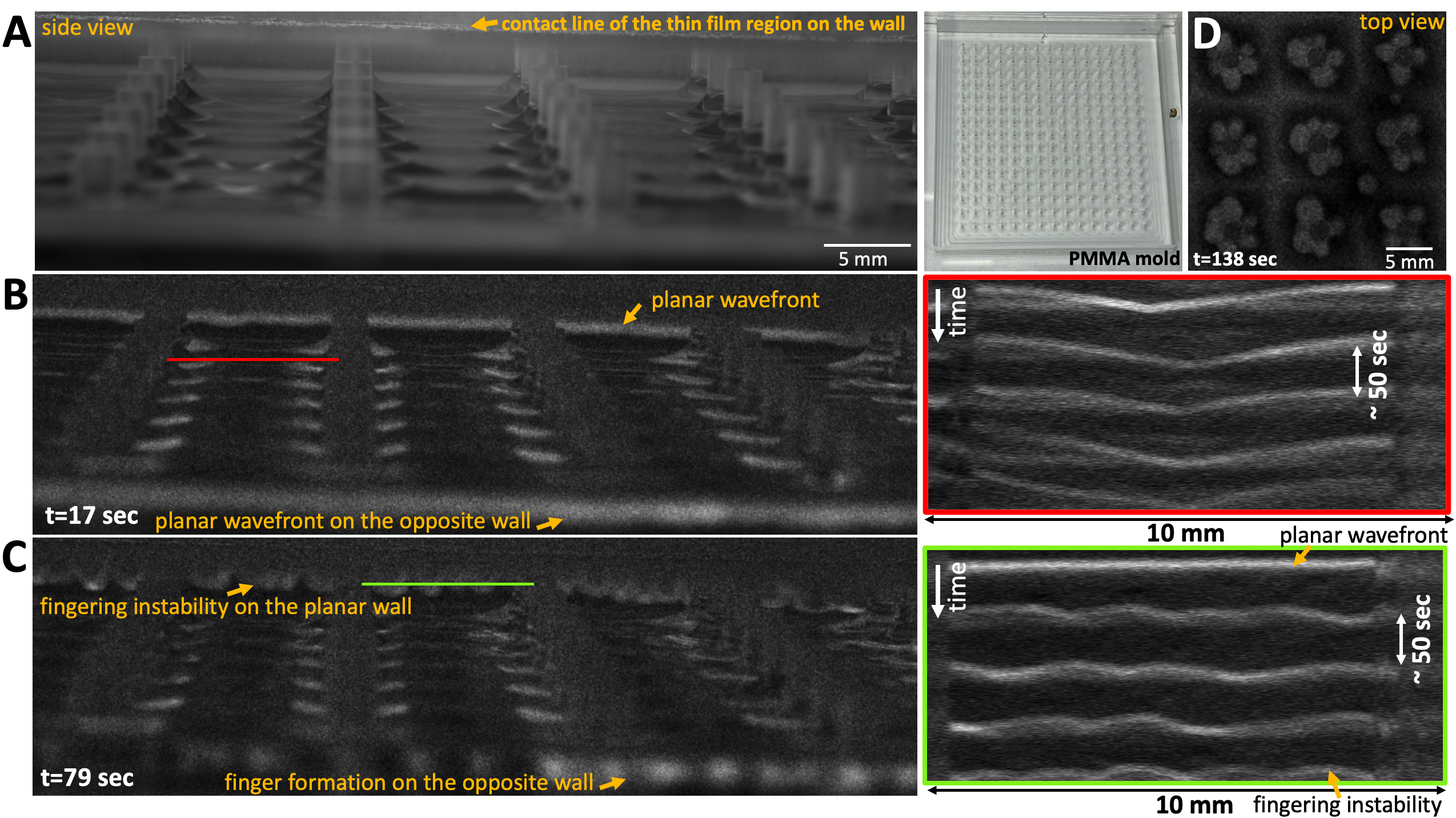}
		\caption{\textbf{Experiments with PMMA-made obstacles}. (\textbf{A}) Side view of the cylindrical obstacle array made from PMMA, with walls forming a square boundary. The fluid climbs up the plasma-treated hydrophilic obstacles as well as the wall boundary. (\textbf{B}) Images are subtracted every 3 seconds to better visualize the wavefronts initiating in the thin film region, wetting the pillars and the walls, and traveling downward. The obstacles act as wave centers, as shown in the space-time plot in the red box, which stacks up light intensity along the red line. (\textbf{C}) Initially, the wavefronts are planar along the wall and circular around the pillars, but they rapidly deform, developing into finger-like patterns. This wavefront instability on the wall is clearly visible in the space-time plot in the green box, created by stacking up the light intensity along the green line on the wall boundary. (\textbf{D}) A top view of the waves around the obstacles, illustrating the breaking of the wavefront and the emergence of flower-like patterns. These images are representative of at least 10 experiments.}
		\label{fig:PMMA}
	\end{center}
\end{figure}

To unravel the mechanisms (i) governing the formation of synchronous circular waves around the pillars and (ii) contributing to the instability of the wavefront in an uncovered setup, we began by examining the hypothesis that the permeability of PDMS to bromine gas (and potentially other chemicals) facilitates wave formation on the pillars, an idea supported by literature~\cite{moustaka2021partition,sheehy2020impact,buskohl2016belousov}. To test this possibility, we conducted experiments using PDMS pillars coated with a thin layer of gold (approximately 200 nm thick, see Figure~\ref{fig:Gold}A). Contrary to expectations, the pillars continued to function as centers for synchronous wave activity even with the gold coating, which blocks chemical absorption (Figure~\ref{fig:Gold}B and Video 8). Additionally, we conducted experiments using obstacles made of acrylic (PMMA) with plasma treatment and observed that these pillars also functioned as centers for wave activity (Figure~\ref{fig:PMMA}A-C). These results confirm that the absorption of chemicals into PDMS is not the primary mechanism driving the formation of concentric waves around the obstacles. 

\begin{figure}[t!]
	\begin{center}
		\includegraphics[width=0.9\columnwidth]{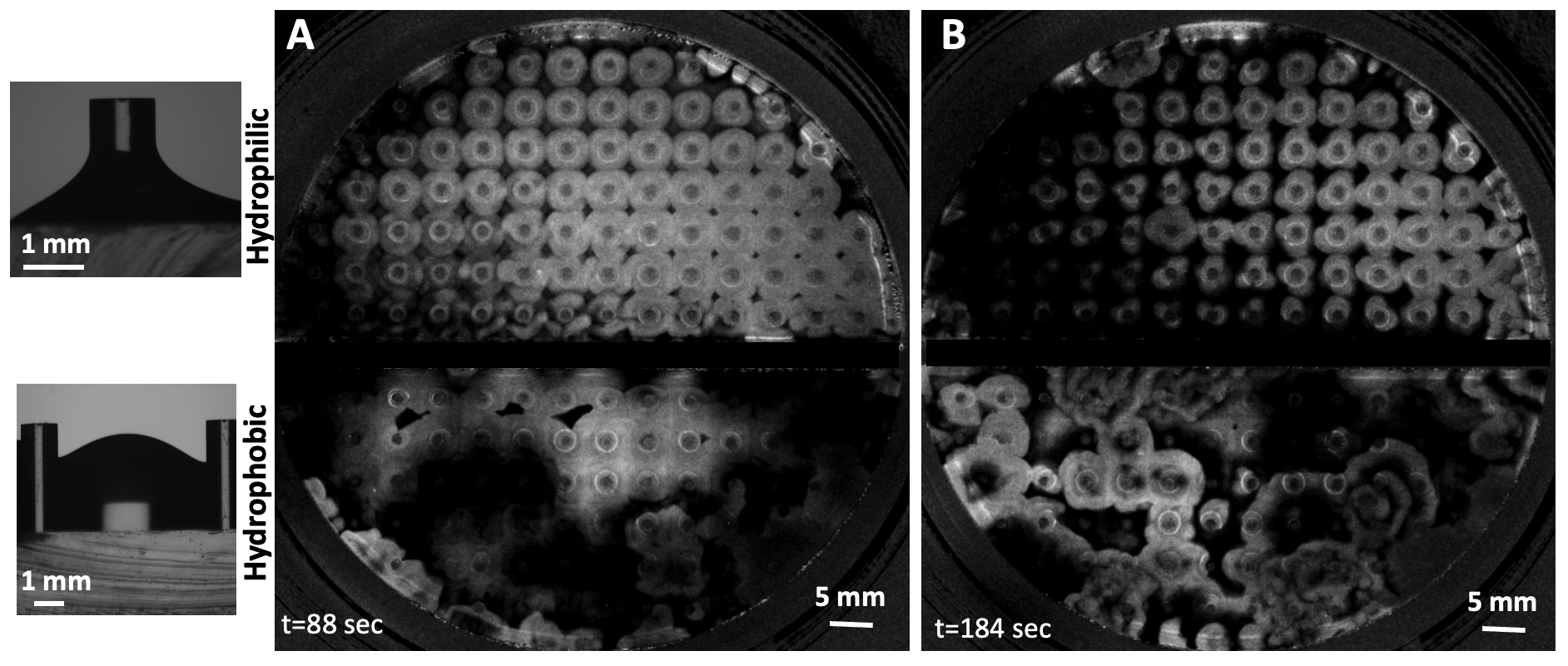}
		\includegraphics[width=0.9\columnwidth]{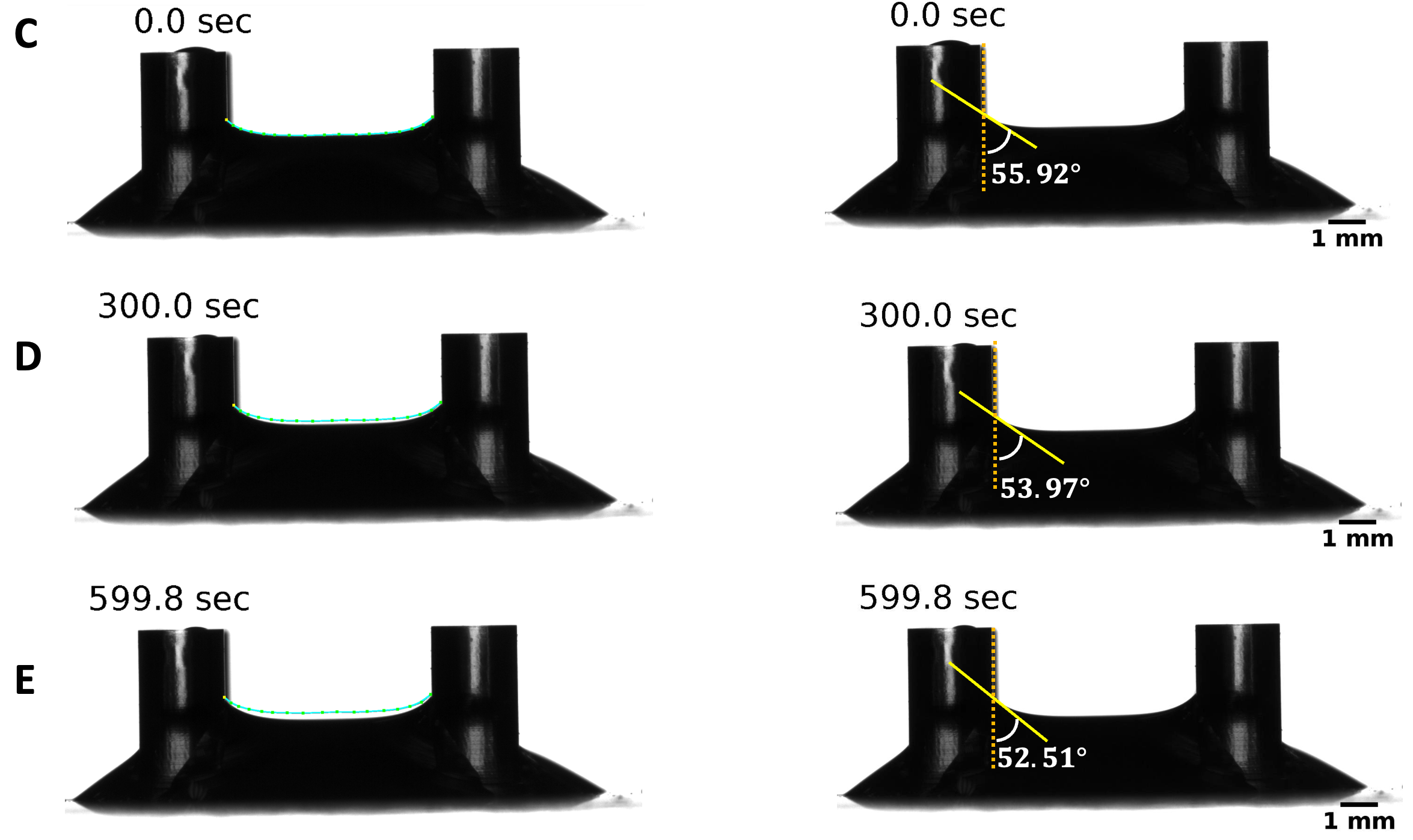}
		\caption{\textbf{Hydrophilic versus hydrophobic obstacles:} An overview of a partitioned, uncovered Petri dish taken at two distinct times: before (\textbf{A}) and after (\textbf{B}) the emergence of instability. The upper portion of the dish has been plasma-treated to render PDMS hydrophilic, whereas the lower section has been left untreated, maintaining its hydrophobic properties (see Video 9). A zoomed-in side view of the fluid near the pillars is also presented, emphasizing the rising fluid around the hydrophilic pillar. (\textbf{C-E}) Side-view imaging of the fluid climbing the hydrophilic pillars confirmed that the contact line remains pinned. As the liquid evaporates, the contact angle gradually decreases (see Video 12). The cyan curves represent the air-liquid interface at $t$$=$0; \color{black}{See the SI for meniscus analysis and curvature calculations}.}
		\label{fig:Partition_New}
	\end{center}
\end{figure}

Next, we explored the impact of the thin liquid layer ascending the PDMS-made pillars, which are hydrophilic due to plasma treatment. Our investigation involved several experimental approaches. Initially, we conducted experiments with pillars that had not undergone plasma treatment and observed that most of the pillars did not act as centers for wave generation (see Figure~\ref{fig:Partition_New}A-B and Video 9). Subsequently, we decreased the height of the pillars from 3 mm to 1 mm, ensuring that these shorter pillars were submerged to the same fluid level. We noted that the submerged pillars also did not act as wave centers (Video 10). 

\begin{figure}[htbp!]
	\begin{center}
		\includegraphics[width=0.7\columnwidth]{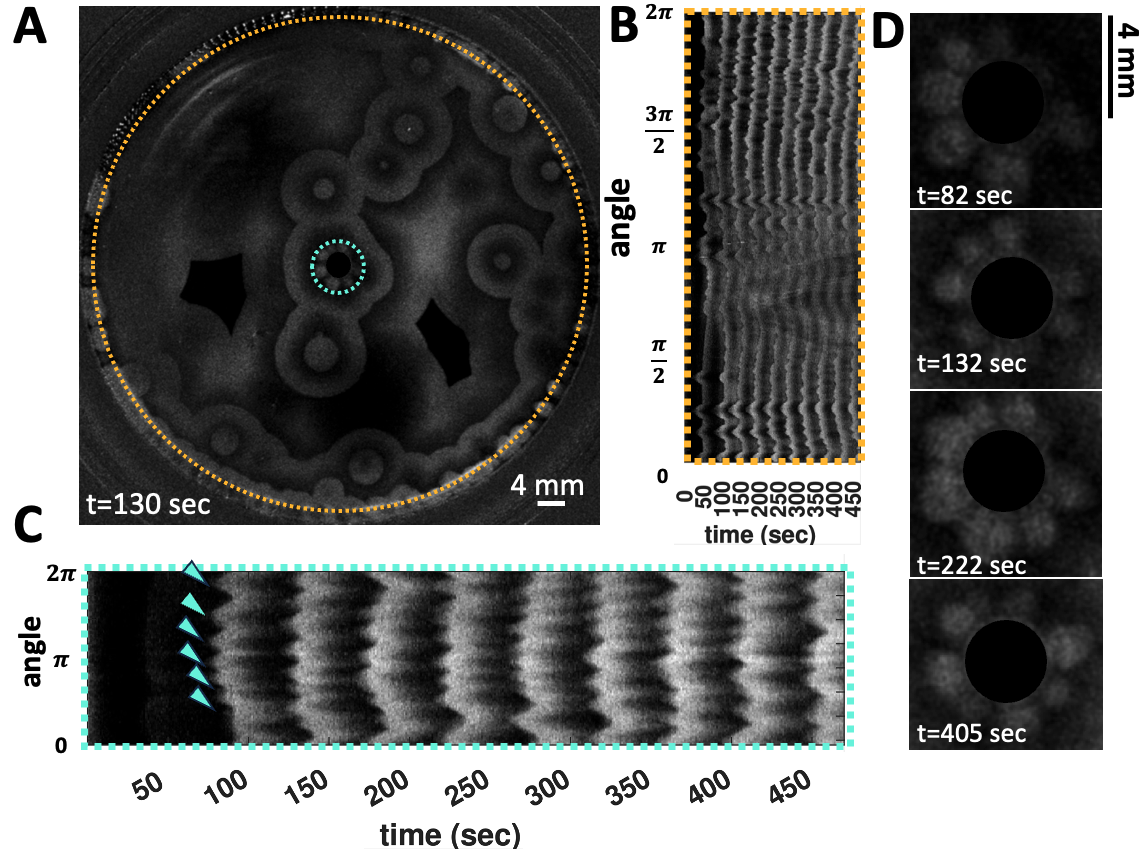}
		\caption{\textbf{A single pillar also acts as the wave center.} (\textbf{A-C}) An isolated pillar serves as the wave center and exhibits wavefront instability in an uncovered Petri dish. Additionally, we observe wavefront instability at the PDMS walls around the periphery. (\textbf{D}) shows a flower-like pattern with six or seven petals at different time points that formed around the solitary central pillar.}
		\label{fig:SinglePillar}
	\end{center}
\end{figure}

Finally, to assess whether the formation of circular waves around the pillars was influenced by the proximity of adjacent pillars, we conducted experiments with a single isolated pillar. As shown in Figure~\ref{fig:SinglePillar} and Video 11, this solitary pillar also acted as a wave center and exhibited wavefront instability in an uncovered setup. Notably, under uncovered conditions, only waves originating from the pillar and the PDMS wall at the periphery exhibited wavefront instability, while off-center waves retained their circular shape without fragmentation. These findings collectively highlight the crucial role of the climbing liquid layer on the pillars in both the generation of synchronous circular waves and the onset of wavefront instability.
\subsubsection{Experimental evidence of a fingering  instability}
\begin{figure}[t!]
	\begin{center}
		\includegraphics[width=0.96\columnwidth]{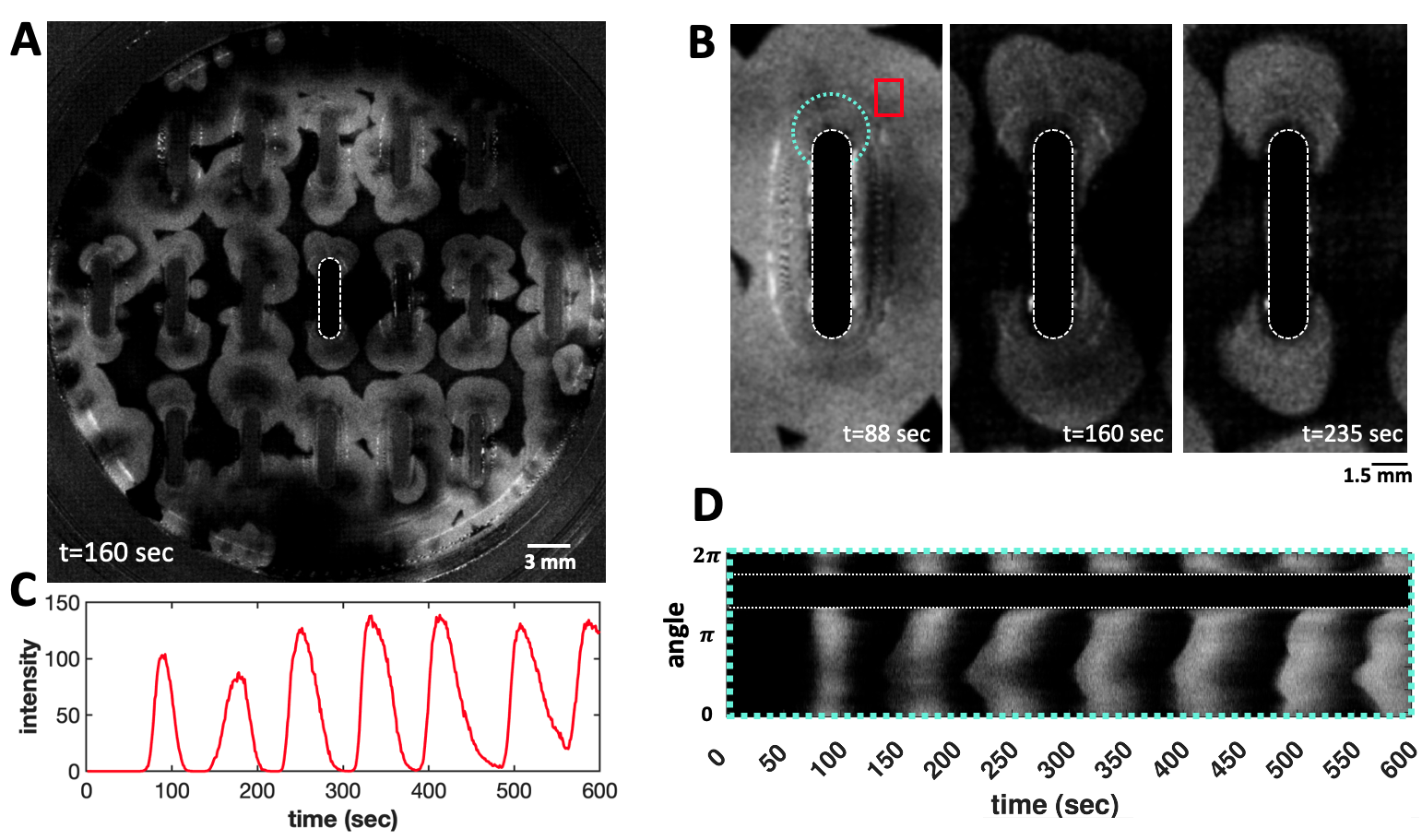}
		\includegraphics[width=0.91\columnwidth]{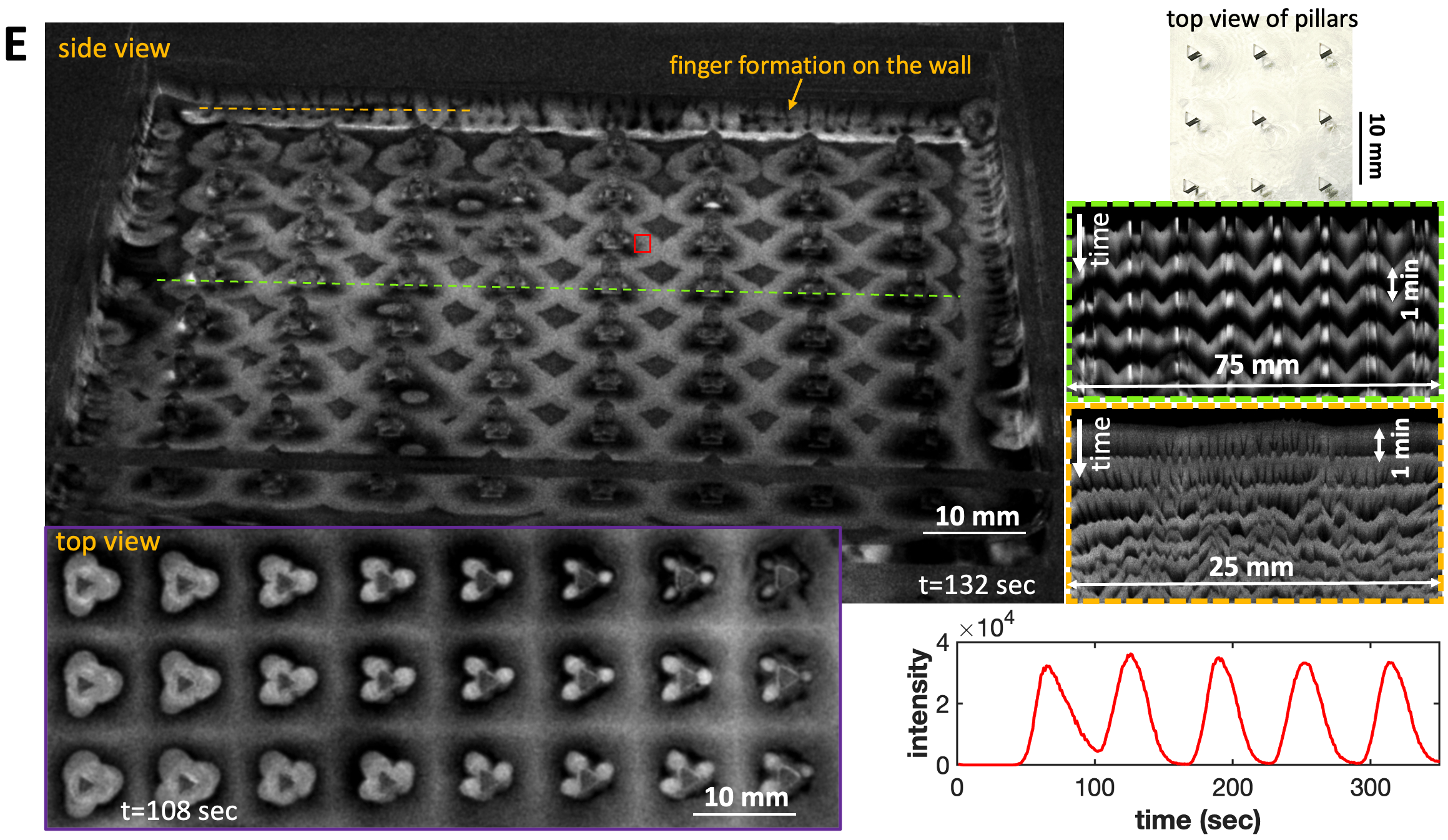}
		\caption{\textbf{Experiments with  rounded rectangular obstacles and triangular obstacles}.  (\textbf{A}) In an experiment examining the impact of evaporation, PDMS-made obstacles with a rectangular body and rounded ends were used. (\textbf{B})  The sequence of wave pattern of the highlighted pillar in panel (A) indicates that instability mainly arises at the curved ends. (\textbf{C}) Oscillations of light intensity averaged over the red box shown in panel (B). (\textbf{D}) A space-time plot along the circular path marked in cyan in panel (B) illustrates the progression of wavefront instability. (\textbf{E}) A side view of an experiment featuring triangular PMMA obstacles within a square boundary, where waves predominantly initiate at the sharp edges. Space-time plots along the green and orange lines are also shown. The intensity averaged over the red box shows oscillations of 60 seconds. Notably, the chemical fingers are clearly visible along the wall.  }
		\label{fig:Rectangular}
	\end{center}
\end{figure}
As previously discussed, our experiments reveal that the climbing film on the external surfaces of cylindrical pillars is essential for the formation of synchronous circular waves and the onset of wavefront instability. We observe that wavefront instability predominantly occurs in uncovered experimental setups, where evaporation is more significant compared to covered setups. Notably, in our control experiments, where the CHD-BZ solution is covered with a thin layer of hexane to suppress evaporation, the obstacles continue to function as wave centers, but no wavefront instability is observed (Figure~\ref{fig:Gold}D-E).

Moreover, side-view imaging of the fluid climbing around the pillars confirms that the contact line remains pinned (Figure~\ref{fig:Partition_New}C-E and Video 12), indicating the presence of upward capillary flows that counteract fluid loss caused by evaporation. Within this ascending liquid layer, the thinner downstream precursor film is expected to evaporate faster than the thicker upstream film, creating slight temperature and/or concentration gradients, due to evaporative cooling and/or selective evaporation of one or more components of the multicomponent liquid. These gradients generate surface tension differences, driving thermal and/or solutal Marangoni flows that ultimately disrupt the wavefront. Our hypothesis that waves originate from the thin ascending fluid layer is supported by our observations in Figure~\ref{fig:PMMA}, using a PMMA mold with square wall boundaries. The side view of the waves confirms that they initiate at the thin film in the contact region, which wets the square wall boundaries, and then travel downward (Video 13). Importantly, in an uncovered setup, the downward-traveling wave along the wall boundary (as well as the waves initiated around the cylindrical obstacles) initially maintain a planar (circular) form but eventually break into fingers at later times (Figure~\ref{fig:PMMA}B-C and D). This fingering instability is similarly visible in Figure~\ref{fig:Rectangular}E and Video 14 with PMMA-based triangular obstacles within square wall boundaries. Lastly, we note that the initiation of waves at the ascending thin film is consistent with a separate experiment in which the Petri dish was slightly tilted~\cite{inomoto1995depth}. In this setup, we observed that synchronization waves initially appeared in regions with lower fluid thickness and subsequently propagated toward areas with higher fluid thickness (see Figure~\ref{fig:Phase}A-B and Video 15). 

Literature reveals that Marangoni-driven thin films on planar or curved substrates exhibit a complex array of dynamics~\cite{smolka2017dynamics, bertozzi1998contact, fanton1996thickness, lohse2020physicochemical}, dependent on the flat upstream film thickness trailing the advancing front. When the upstream film thickness is set by the competition effect of meniscus curvature and the surface tension gradient,  the film is sufficiently thin for the advancing front to form a stationary compressive or Lax shock, which are well known to be unstable to contact line perturbations~\cite{fanton1996thickness,bertozzi1998contact,smolka2017dynamics,bertozzi1999undercompressive}. It is important to note that surface tension gradients, which drive Marangoni flows, can result from temperature gradients, concentration gradients, or a combination of both. Concentration gradients often develop due to the evaporation of more volatile chemical components within the ascending thin film, similar to the mechanism observed in the tears of wine phenomenon~\cite{fournier1992tears, hosoi2001evaporative}. The interplay between Marangoni flows and gravity can disrupt the contact lines, evolving into a finger pattern. Specifically, the linear stability analysis in Ref.~\cite{smolka2017dynamics} correlates linearly the number of fingers with the cylindrical pillar's perimeter, aligning with our findings (see Figure~\ref{fig:VaryingDiameter}). This analysis also suggests a minimum pillar diameter threshold below which fingering instability does not occur. To validate this, we reduced the pillar diameter to 0.5 mm in our uncovered setup experiments, observing that the wavefronts remained nearly circular (see Figure~\ref{fig:VaryingDiameter}E and Figure~\ref{fig:ThinPillars}). This resembles the situation of Rayleigh-Bénard convection where, for given driving strength, the linear instability cannot develop for containers with a too small aspect ratio (width/height) \cite{shishkina2021rayleigh}.

\begin{figure}[t!] 
	\begin{center}
		\includegraphics[width=\columnwidth]{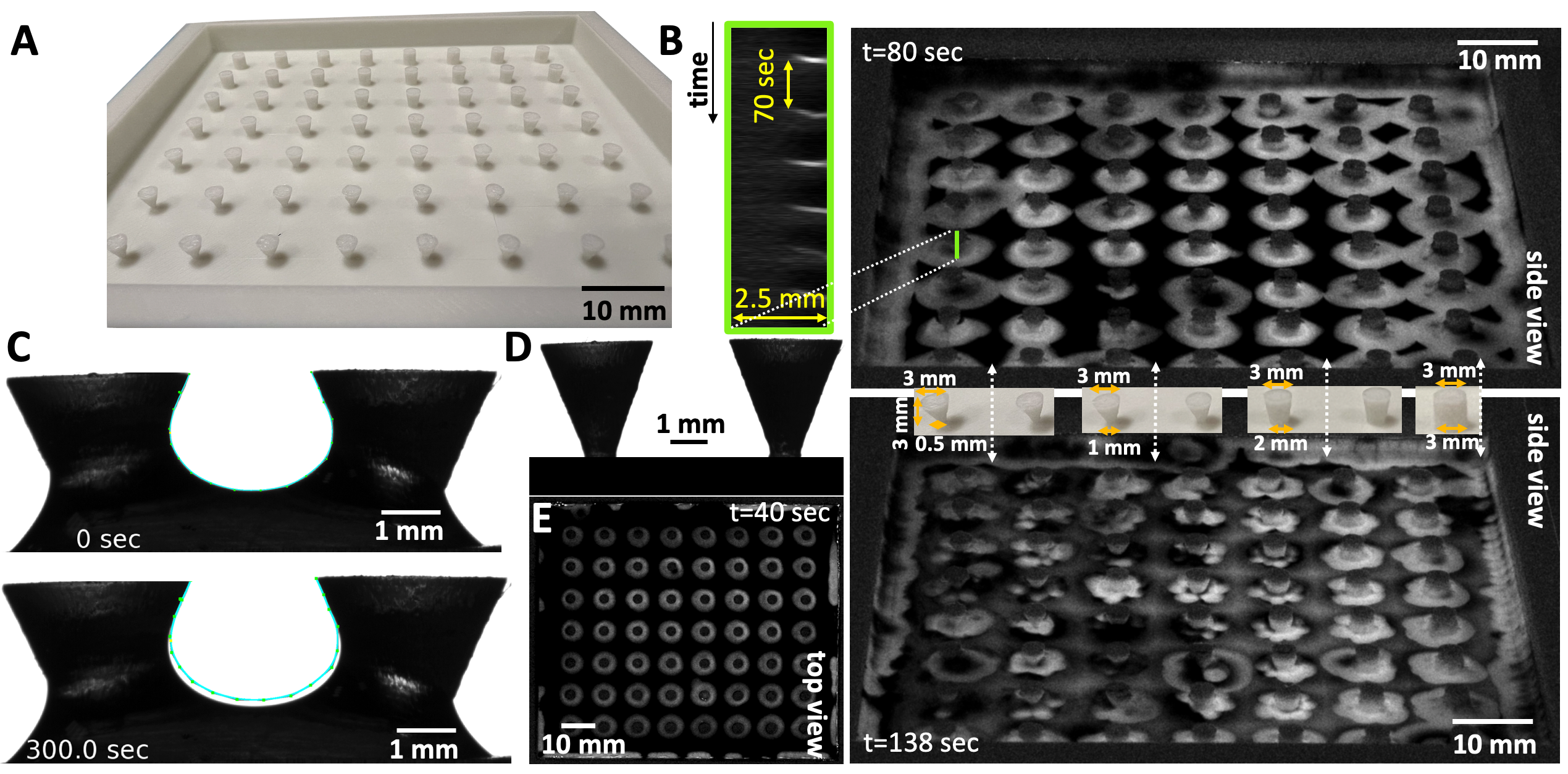}
		\includegraphics[width=\columnwidth]{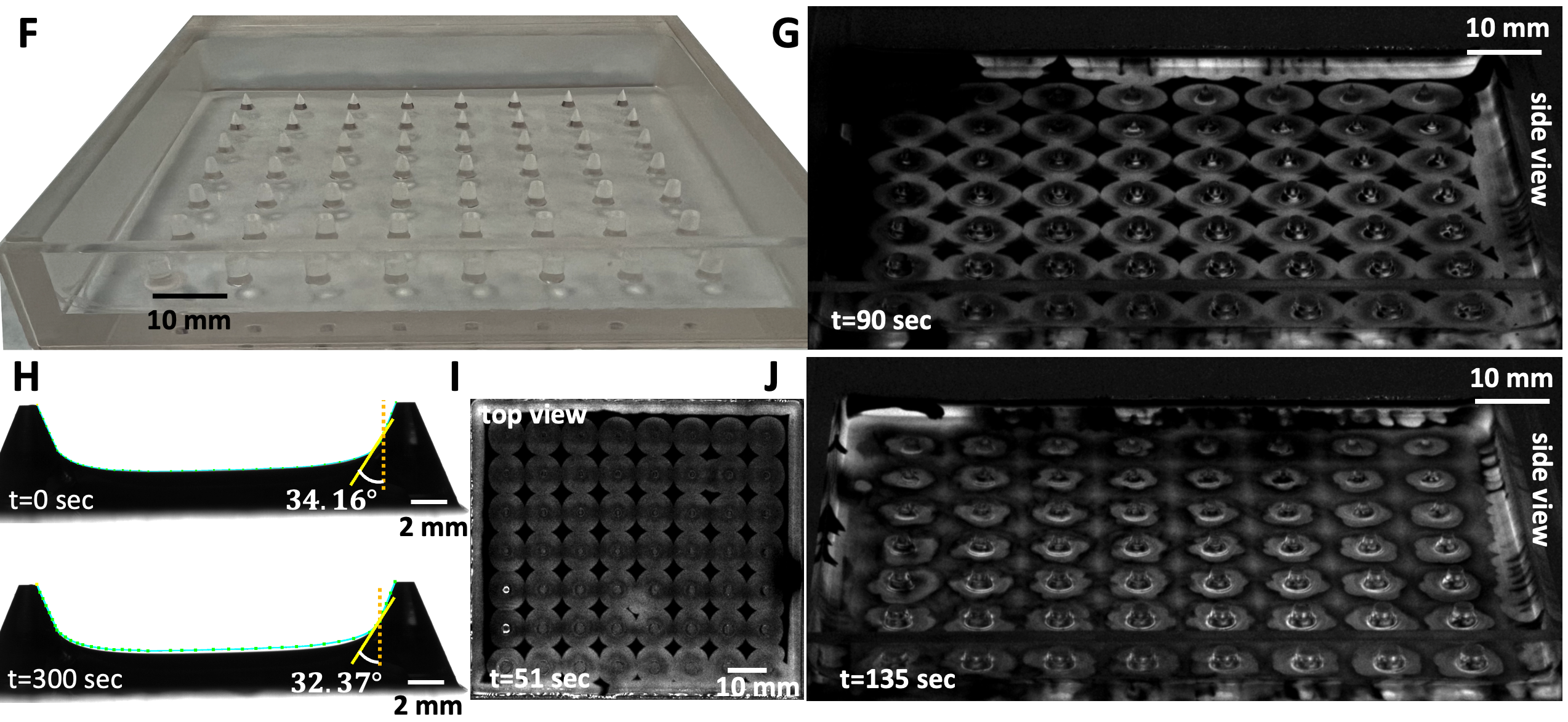}
		\caption{\textbf{Experiments with both inverted tapered and tapered pillars.} (\textbf{A}) Conical pillars with base diameters ranging from 0.5 mm to 2 mm, with the top row displaying standard cylindrical pillars of 3 mm diameter. (\textbf{B}) Side view of centered waves observed before and after the onset of instability. The space-time within the green box represents the light intensity along the green line over time, revealing that the waves originate from the climbing thin film and propagate downward. (\textbf{C-D}) A close-up side view of two neighboring cones with a base diameter of 1 mm. The cyan curve illustrates the air-liquid interface at $t$$=$0. (\textbf{E}) Top view of the centered waves prior to the instability. (\textbf{F}) Tapered pillars with a 3 mm base diameter that gradually decrease to smaller diameters toward the top. (\textbf{G-J}) The side and top views of the centered waves are shown, along with the contact angle of the CHD-BZ solution around the hydrophilic conical structures at $t$$=$0. Over time, as evaporation progresses, the contact angle decreases ($t$$=$300 sec). The cyan curves in panel (H) represent the liquid-air interface at $t$$=$0; see Videos 17-20.}
		\label{fig:Cones}
	\end{center}
\end{figure}

\begin{figure}[t!]
	\begin{center}
		\includegraphics[width=0.95\columnwidth]{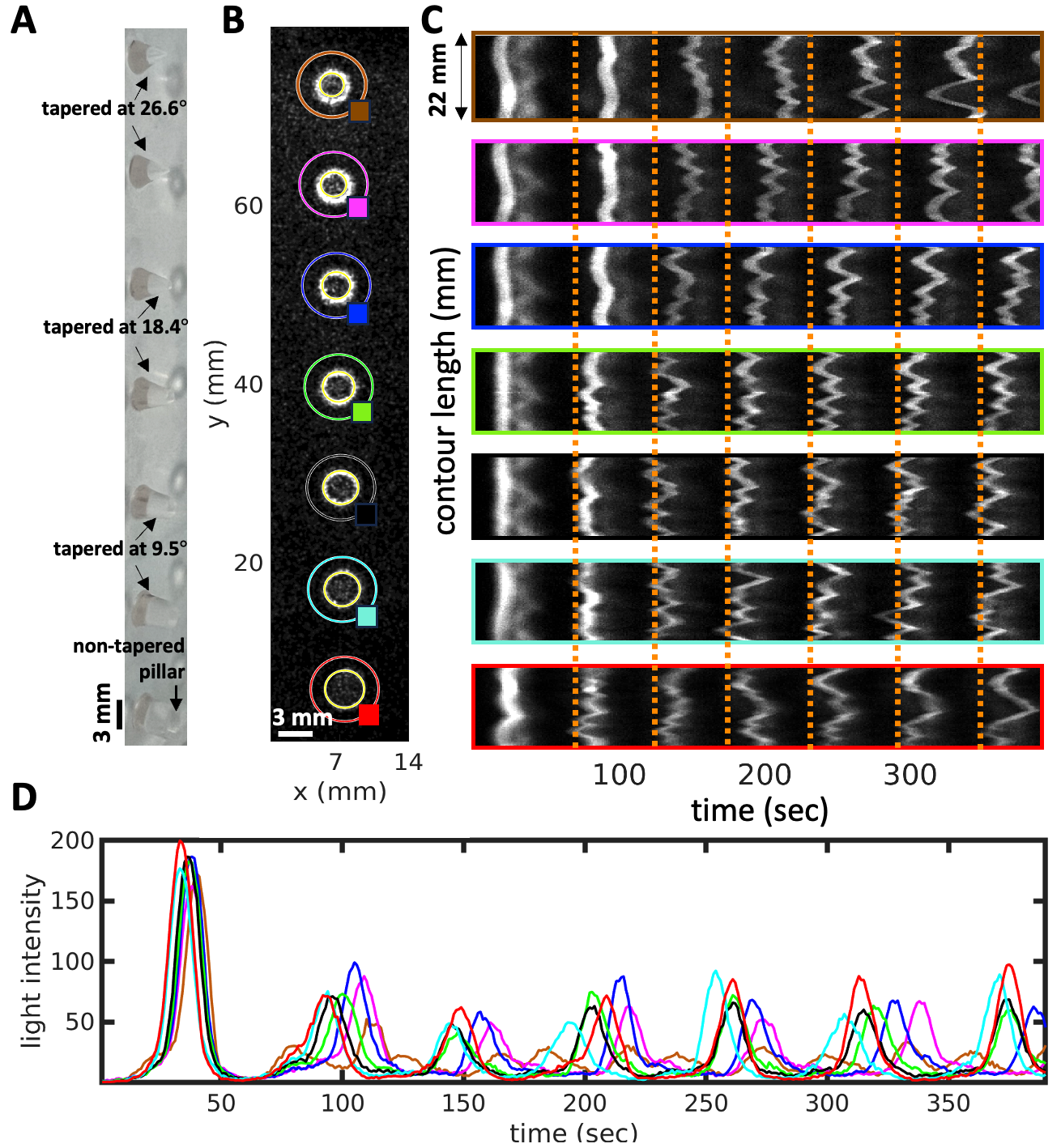}
		\caption{(\textbf{A)} Side-view of tapered pillars used for data processing of waves centered around them in panels (B–D), with the last pillar at the bottom representing the standard non-tapered pillar. (\textbf{B-D}) The top view highlights the circles along which the space-time plot in panel (C) is generated. The colored squares indicate the areas where light intensity is averaged and plotted over time in panel (D). Both the space-time plot and light intensity oscillations show that after the first cycle of nearly synchronous circular waves, the second round of waves with a broken wavefront first appears in the non-tapered pillar at the bottom, followed by a time delay in the tapered pillars. The two tapered pillars at the top, both tapered at approximately 26$^\circ$, exhibit the longest delay. The orange lines serve as visual guides (see Video 18).}
		\label{fig:ConesOneRow}
	\end{center}
\end{figure}

Moreover, we tested the impact of evaporation on fingering instability using obstacles with a rectangular geometry and circular ends (see Figure~\ref{fig:Rectangular}A-D and Video 16). Given the increased surface area at the curved boundaries, we anticipated a higher evaporation rate here, which likely triggers the fingering instability predominantly at these locations, a hypothesis supported by our observations. A similar pattern was noted in our experiments with obstacles of triangular cross-section, where waves primarily started at the sharp boundaries (Figure~\ref{fig:Rectangular}E and Video 14). Additionally, we conducted experiments with both inverted tapered and tapered pillars, confirming that as long as a thin wetting layer surrounds the obstacles, they act as synchronous wave centers (Figure~\ref{fig:Cones} and Videos 17-20). In the experiments with tapered pillars, we observed that wavefront instability first appeared in non-tapered pillars (our standard straight pillars with a 0$^\circ$ taper angle) and, after a time delay, in the tapered pillars as the angle gradually increased from 0$^\circ$ to approximately 26$^\circ$ (Figure~\ref{fig:ConesOneRow}). We attribute this effect to the higher evaporation rate in standard pillars compared to tapered pillars, due to their greater available surface area at the fluid-pillar contact line. Interestingly, we found that the average number of petals remains unchanged across different taper angles. However, as the perimeter of the pillars at the fluid-pillar contact line decreases with larger taper angles (see yellow circles in Figure~\ref{fig:ConesOneRow}B), the wavelength also decreases as the taper angle increases, ensuring that the same number of petals is accommodated.

Furthermore, we observed that during multiple cycles of covering and uncovering the setup, the wavefronts, which exhibit flower-like patterns when uncovered, revert to their circular form once the setup is covered, and vice versa (Videos 21-22). To further investigate the role of boundaries, we conducted a control experiment using an uncovered, plasma-treated glass Petri dish without any obstacles (Figure~\ref{fig:Glass} and Video 23). Interestingly, we observed wavefront instability at the glass boundary, closely resembling the phenomena observed at the PDMS wall in Figure~\ref{fig:PetriOpen_SmallDish} and the PMMA wall in Figures~\ref{fig:PMMA} and \ref{fig:Rectangular}E. A similar control experiment was performed in an uncovered Petri dish with a PDMS substrate (no pillars) and a PDMS wall boundary, where the wavefront instability was only observed at the PDMS wall (see Video 24). Overall, these experiments confirm that the thin layer of climbing fluid at the boundary is critical for initiating the waves, and that evaporation significantly contributes to the wavefront instability. 

\begin{figure}[t!]
	\begin{center}
		\includegraphics[width=\columnwidth]{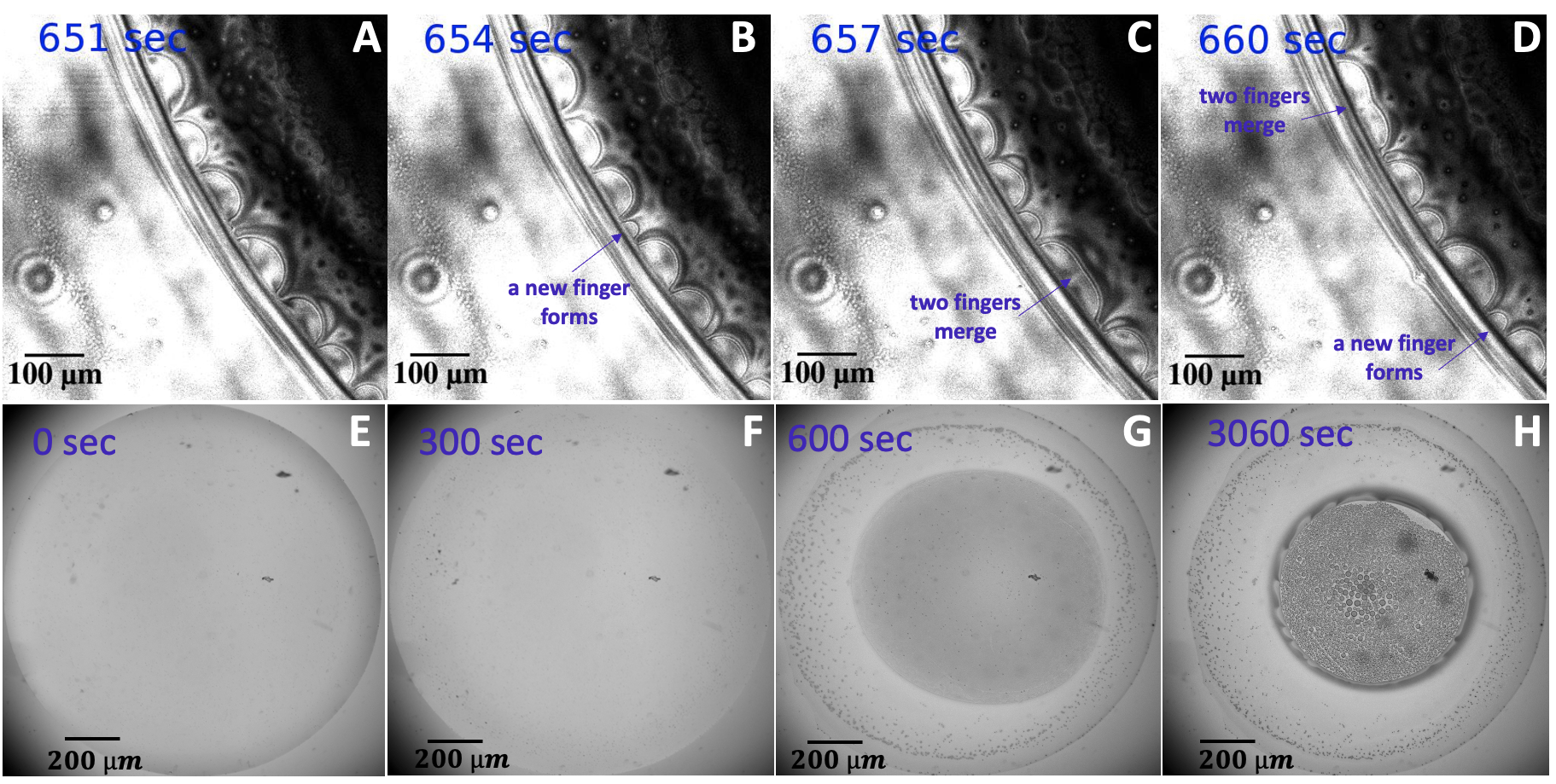}
		\caption{\textbf{Symmetry breaking in an evaporating CHD-BZ droplet and coffee stain effect}. (\textbf{A-D}) A droplet of volume $2.5~\mu l$ is deposited on a hydrophobic glass substrate.  After about 500 sec, the azimuthal symmetry breaks and spots form at the rim evolving over time ($T$$=$22$^\circ$C, RH$=$$40\%$, $\theta_0$$=$12$^\circ$). The images are captured in the reflection mode of a confocal microscope; see Video 25.  (\textbf{E-H}) In a separate experiment imaging the entire droplet, micron-sized beads were introduced to track the flow. During the first 5 minutes of evaporation, the droplet boundary remains pinned to the substrate, leading to the accumulation of tracer particles at the rim due to capillary flows, clearly exhibiting the coffee-stain effect, as depicted in panels (F-H). As evaporation continues and the reaction progresses, a secondary instability develops at the rim of the shrinking droplet, as shown in panel (H) (see Video 26). }
		\label{fig:Droplet}
	\end{center}
\end{figure}

Our final experiment that clearly proves the important role of evaporation in spontaneous breaking of azimuthal-symmetry is an evaporating CHD-BZ droplet. The preferential evaporation of more volatile chemical components at the rim can contribute to the segregation of other chemicals near the rim and eventually break the azimuthal symmetry. Figure~\ref{fig:Droplet}A-D and Video 25 show spot formation near the boundary of a droplet. A similar effect has been reported in water/1,2-hexanediol binary drops in Refs.~\cite{diddens2024non,li2018evaporation,li2020rayleigh,li2022physicochemical}, where 1,2-hexanediol micro-droplets nucleate at the rim of the droplet. We note that in our evaporating droplet experiments, where micron-sized beads were added as tracer particles to track the flow, we observed the coffee-stain effect (see Video 26 and Figure~\ref{fig:Droplet}E-H.). This effect is characterized by the formation of a ring-like structure of deposited beads after the droplet evaporates. The phenomenon arises from a combination of the non-uniform evaporation rate along the droplet interface and the presence of a pinned contact line, which generates a continuous capillary flow from the center of the droplet toward the contact line to compensate for liquid loss during evaporation~\cite{van2022competition,diddens2021competing}. Furthermore, we identified a secondary instability at the droplet boundary (Figure~\ref{fig:Droplet}H), which requires further investigation in future studies.
\subsection{Estimation of dimensionless numbers and timescale comparison}
In our experiments, the chemical fingers forming on the pillars or on the walls of the PDMS/PMMA setups exhibit a wavelength of approximately $\lambda\approx 2~mm$, see Figure \ref{fig:VaryingDiameter}F, inset. We hypothesize that the fluid fingers have a similar wavelength. Assuming the climbing fluid depth is in the range $h=20-100~\mu$m~\cite{vuilleumier1995tears,hosoi2001evaporative}, we calculate the Marangoni stress $\tau$ from the balance with gravity,  $\tau=2 \rho g h /3$~\cite{fanton1996thickness}, which represents the balance between the competing Marangoni stress and gravitational forces. This calculation yields a surface stress range of  $\tau=0.13-0.67$ $N/m^2$. Evaluating the corresponding solutal Marangoni number for $\tau=0.67$ $N/m^2$ and $h=20~\mu m$ yields:
\[\mathscr{M}_S=\frac{\tau h^2}{\mu D}\sim\frac{0.67N m^{-2}\times(20\times10^{-6})^2m^2}{10^{-3} N s m^{-2} \times1.2\times10^{-9}m^2 s^{-1}}\sim 250,\]
where $\mu$ is the viscosity of the CHD-BZ solution and $D$ is molecular diffusivity of chemical components. The corresponding thermal Marangoni number, defined as:\[\mathscr{M}_T=\frac{\tau h^2}{\mu \kappa},\]
is two orders of magnitude smaller due to the ratio of molecular diffusivity to thermal diffusivity, given by the inverse Lewis number\[\ \mathrm{Le}^{-1}=D/\kappa=(1.2\times10^{-9} m^2 s^{-1})/(1.5\times10^{-7} m^2s^{-1})\sim10^{-2},\] 

Indeed, in general, for evaporating multicomponent liquids with different volatilities of their components, solutal Marangoni effects are more relevant than thermal ones \cite{diddens2017evaporating, rocha2024marangoni}.
We now proceed by considering a dimensional analysis similar to Ref.~\cite{hosoi2001evaporative} that examines the interaction between a deformable free surface and the steady-state concentration profile, which is governed by diffusion, evaporation, and convective transport along the pillars. Evaporation reduces the concentration of the more volatile chemical component with lower surface tension near the surface; however, under the assumption of a quasi-steady state, fresh fluid from the reservoir continuously replenishes the depleted region, establishing a stable vertical concentration gradient. We note that given the complexity of the CHD-BZ reaction, where multiple chemical components are continuously produced and consumed, the validity of this assumption requires further extensive investigation in future studies. If the free surface experiences an upward perturbation, the concentration profile must adjust before the interface returns to its original shape. If the concentration realigns more rapidly than the interface relaxes, the perturbed region develops a lower concentration than its surroundings, generating a surface tension gradient that draws fluid into the bump and amplifies the perturbation. Similarly, depressions in the free surface create local surface tension minima, pulling fluid away and reinforcing the deformation. This feedback loop sustains convection and enhances surface fluctuations, making the film susceptible to instability driven by Marangoni forces. The system's stability depends on the relative timescales of diffusion and interface relaxation. If the diffusive timescale is significantly shorter than the relaxation time, the system is unstable, i.e., $T_{\mathrm{diff}}/T_{\mathrm{relax}} \ll 1.$ The relaxation time \( T_{\mathrm{relax}} \) is estimated by balancing pressure gradients with viscous stresses, 
\[	\nabla p \sim \left( \frac{\mu}{h^2} \right) V_f \sim \left( \frac{\mu}{h^2} \right) \frac{\lambda}{T_{relax}},\]
where $V_f\sim0.5$ \(mm/s\) is the estimated speed of fluid movement up the pillars. The restoring pressure gradient is determined by either surface tension or gravity:
\[	p \sim \frac{\gamma h}{\lambda^2} \quad {\mathrm{or}} \quad p \sim \rho g h.\]

For $h=20~\mu m$ and surface tension $\gamma=65~mN/m$~\cite{ukitsu1993generation}, the corresponding timescale comparison yields the dimensionless ratio for surface tension restoring pressure as: \[\frac{T_{\mathrm{diff}}}{T^\gamma_{\mathrm{relax}}}=\left(\frac{h}{\lambda}\right)^4\frac{\gamma h}{D\mu}\sim\left(\frac{20}{2000}\right)^4\frac{0.065\times20\times10^{-6}}{10^{-9}\times10^{-3}} \sim 10^{-3},\] while for gravitational restoring pressure, it is \[\frac{T_{\mathrm{diff}}}{T^g_{ \mathrm{relax}}} =\left(\frac{h}{\lambda}\right)^2\frac{h^3\rho g}{D\mu}\sim\left(\frac{20}{2000}\right)^2\frac{(20\times10^{-6})^3\times1000\times10}{10^{-9}\times10^{-3}}\sim 10^{-2}.\] Since both values are significantly less than one, we conclude that the system is susceptible to Marangoni-driven convection, where surface tension gradients enhance surface deformations and sustain instability. We note that accurately estimating the characteristic timescales and length scales in our study requires simultaneous experimental measurements of film thickness and concentration gradients driving Marangoni flows. Such simultaneous measurements are experimentally challenging, and efforts to address this are currently ongoing in our laboratory.
\subsection{Numerical Simulations}
{\color{black} Comprehensive numerical modeling of the CHD–BZ reaction—including its full chemical kinetics—is beyond the scope of this manuscript and will be pursued in future work. To capture the Marangoni-driven hydrodynamic instability qualitatively, we instead employ a simplified two-component fluid mixture exhibiting Marangoni instabilities -- e.g., a binary ethanol–water mixture, in which ethanol is both more volatile and has lower surface tension than water.  We then reproduce the experimental geometry by placing this mixture around a pillar (Fig. \ref{fig:geometry}) and considering diffusion-limited evaporation of both components. The contact line remains pinned to the pillar walls, and a symmetry boundary condition is imposed on the right. Our evaporation model follows the approach of Section 2.1 in Ref. \cite{rocha2024marangoni}, suitably adapted for this configuration (see Method Section~\ref{Method}).
\begin{figure}[htbp!]
	\centering
	\includegraphics[width=1\columnwidth]{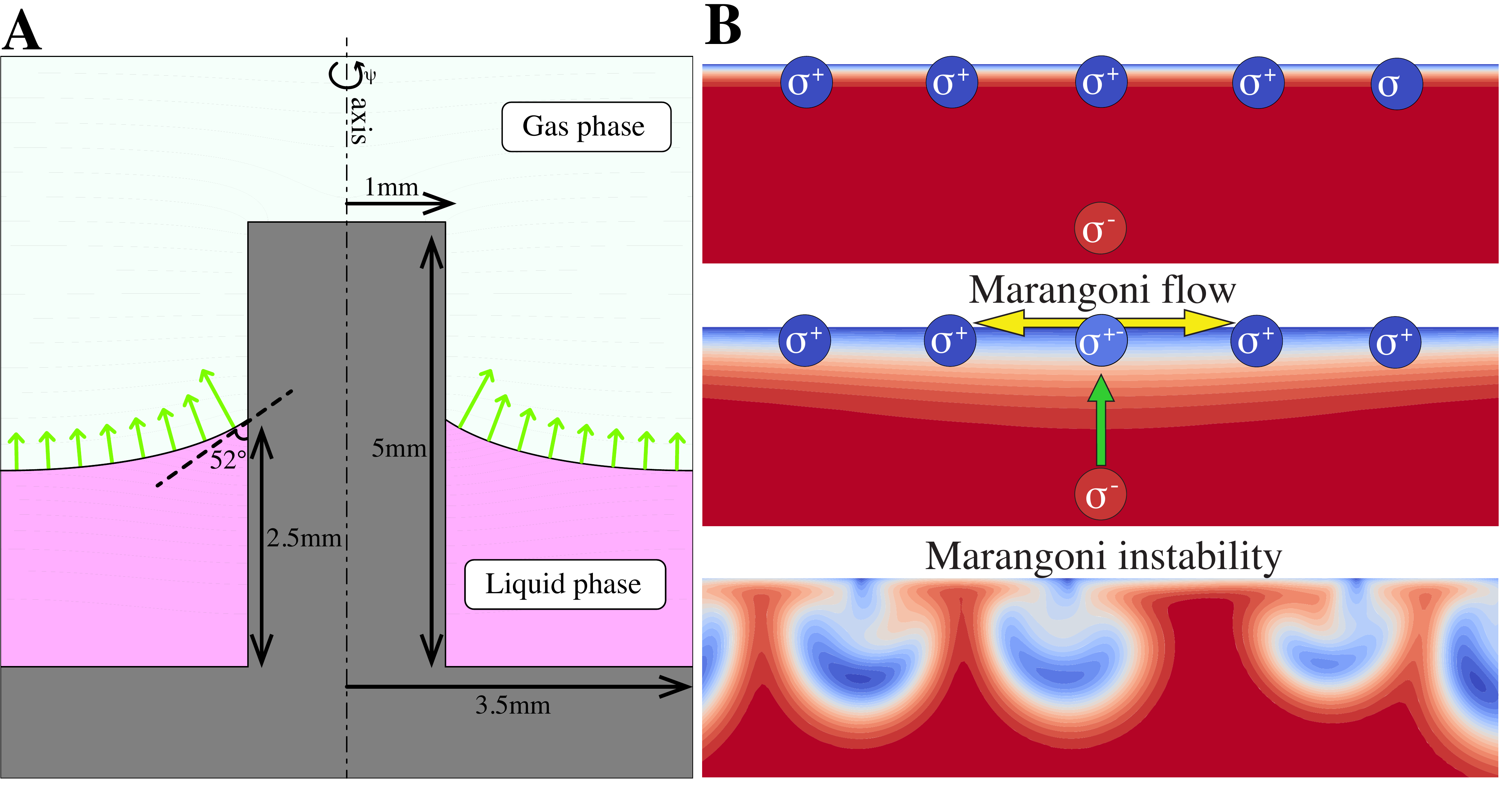}
	\caption{{\color{black}{(\textbf{A}) Geometry of the problem setup based on the experiments at time $t$=599 s (see Fig.~\ref{fig:Partition_New}E). The liquid layer (magenta) is pinned to the pillar walls, forming an angle $\theta$ = 52$^\circ$ with the pillar walls and it is at a height of \SI{2.5}{\mm} at the walls. The liquid evaporates through diffusion of vapor in the gas phase, as schematically represented by the green arrows. Note the longer arrows close to the contact line, representing the larger evaporation flux close to the contact line singularity. (\textbf{B}) Schematics of the Marangoni instability phenomena in a box with periodic side boundary and an evaporating top surface, where only one of the two components evaporates. When the evaporating component has a lower surface tension, the interface increases its concentration on higher surface tension fluid. May a perturbation locally reduce the surface tension somewhere in the interface, it will onset a Marangoni flow, which in turn draws fluid away from the perturbed region (yellow arrows in second panel). Due to continuity, the displaced fluid must be replenished, which brings fluid from the bulk towards the perturbed region (green arrow in second panel). This draws lower surface tension fluid to the perturbed region, further decreasing the surface tension and creating a positive feedback loop that destabilizes the system -- leading to the so-called Marangoni instability.
    }}}
	\label{fig:geometry}
\end{figure}
%

First, we compute the axisymmetric quasi-stationary base state as a function of the Marangoni number, $\mathrm{Ma}$. Next, we carry out a linear stability analysis with respect to azimuthal perturbations of wavenumber $m$. Our results reveal that the radial width of the liquid layer strongly influences stability: wider layers support longer-wavelength modes. As illustrated schematically in Figure\ \ref{fig:geometry}B (see also Figure\ 1 of Ref.\ \cite{rocha2024marangoni}), a local drop in surface tension drives Marangoni flow away from the perturbed region. The resulting pressure gradient pulls fluid -- enriched in the low-surface tension component -- from the bulk back into that region, creating a positive feedback loop that can trigger the instability. The resulting positive feedback can destabilize the system, leading to a Marangoni instability. 
A wider liquid layer reduces flow restrictions, thereby allowing longer-wavelength perturbations to develop. 

To specifically assess the role of the Marangoni number in this confined geometry and avoid varying geometry effects, the liquid width is fixed at  3.5 mm. We then vary $\mathrm{Ma}$ and analyze the stability of the base compositional state with respect to perturbations of different azimuthal wavenumbers. Figure~\ref{fig:dispersion_relation} shows the real eigenvalues of the system as functions of both $m$ and $\mathrm{Ma}$. 
The results reveal that the system remains stable for low $\mathrm{Ma}$ values ($\mathrm{Ma} \lessapprox 100$) but becomes unstable as $\mathrm{Ma}$ increases. At low $\mathrm{Ma}$, the dominant instabilities arise from perturbations with lower $m$, while higher $\mathrm{Ma}$ shifts the instability toward larger $m$. 
\begin{figure}[t!]
	\centering
	\includegraphics[width=0.5\columnwidth]{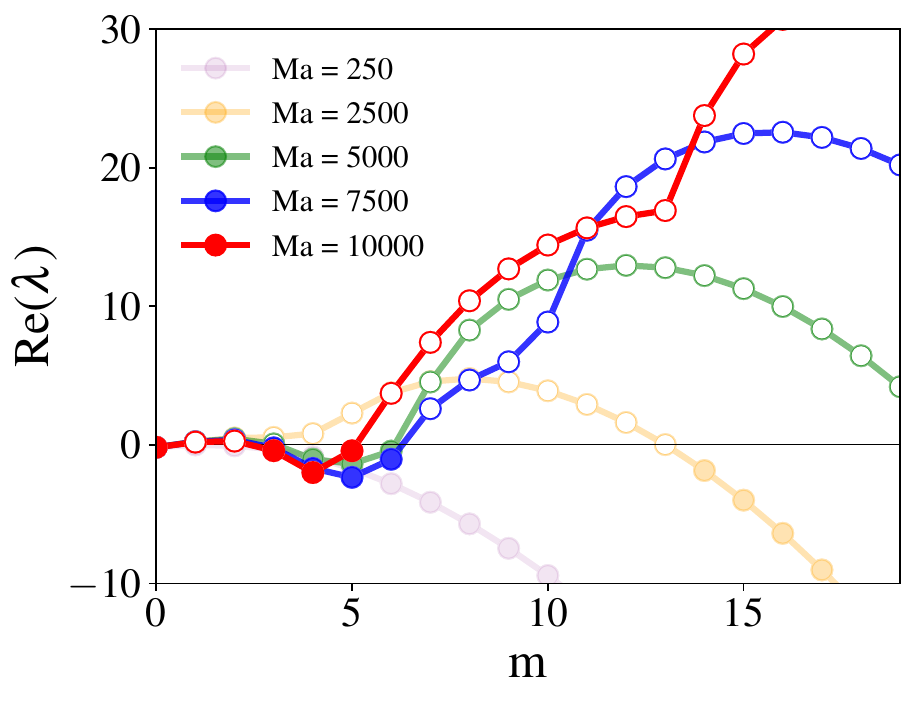}
	\caption{{\color{black}{Dispersion relation of the system for different $\mathrm{Ma}$ values.
		The real part of the nondimensionalized eigenvalues is shown as a function of the azimuthal wavenumber $m$ for $\mathrm{Ma}$ values of 100, 2500, 5000, 7500, and 10000.
		The $\mathrm{Re}(\lambda) = 0^-$ line indicates the stability threshold, where the real part of the eigenvalue becomes positive, indicating instability.}}}
	\label{fig:dispersion_relation}
\end{figure}

To visualize a fully three‐dimensional compositional field, we can plot
\[
y(r,z,\psi) \;=\; y^0(r,z)\;+\;\epsilon\,y^m(r,z)\,e^{i m \psi}\quad + \mathrm{c.c.},
\]
where \(\epsilon\) is a small perturbation amplitude, \(y^0\) is the axisymmetric base state, and \(y^m\) is the eigenfunction for azimuthal mode \(m\), see Figure~\ref{fig:eigenfunctions} showing the full flow (panel A) and the mode $m=5$ (panels B-D). This representation assumes a single dominant eigenmode -- which may not strictly hold here -- but serves to qualitatively illustrate the emerging instability pattern.
\begin{figure}[t!]
	\centering
	\includegraphics[width=0.75\columnwidth]{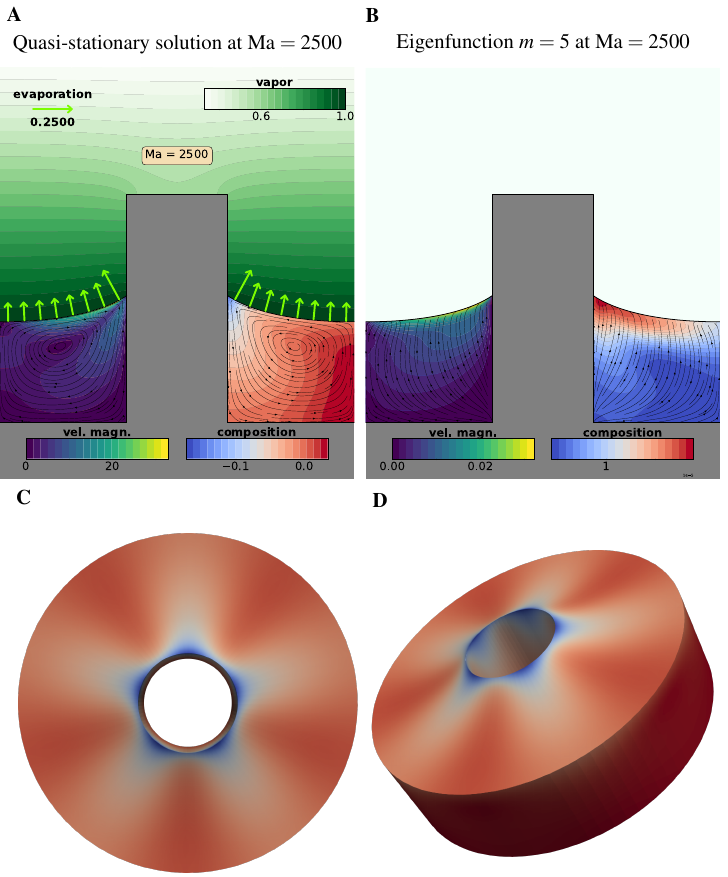}
	\caption{{\color{black}{(\textbf{A}) Nondimensional axisymmetric base solution at $\mathrm{Ma} = 2500$ for liquid's velocity magnitude (left), liquid's composition (right) and gas vapor's concentration. In the liquid-gas interface, the arrows shown represent the evaporation rate. (\textbf{B}) Eigenfunction with azimuthal mode $m=5$ for the liquid's $(r,z)$-velocity magnitude (left) and liquid's composition (right). (\textbf{C}) Top view of the three-dimensional expansion $y(r,z,\psi) = y^0(r,z) + \epsilon y^m(r,z) e^{i m \psi} + \mathrm{c.c.}$ for the composition. (\textbf{D}) Isometric view of the same expansion.
 }}}
	\label{fig:eigenfunctions}
\end{figure}}
\section{Conclusions and outlook}
In this study, we introduce a periodic array of hydrophilic cylindrical obstacles into the CHD-BZ reaction, which rapidly become synchronized centers of wave activity. The thin layer of chemical solution ascending the obstacle surfaces is essential for enabling these structures to function as wave sources, thus precisely determining their spatial positions (Figure~\ref{fig:PetriCloseOpen}C). Although an isolated obstacle can also act as a wave center (Figure~\ref{fig:SinglePillar}), an array arrangement effectively prevents the formation of random wave centers elsewhere in the solution. Under covered conditions, where evaporation is minimized, wavefronts maintain stable circular shapes through multiple cycles. In contrast, uncovered conditions—with significant evaporation—induce Marangoni-driven instabilities, causing wavefront breakage and the emergence of striking flower-like patterns (Figure~\ref{fig:PetriCloseOpen}F). The number of petals in these patterns scales linearly with obstacle diameter (Figure~\ref{fig:VaryingDiameter}), and there is a minimum diameter required for this instability to manifest (Figure~\ref{fig:ThinPillars}). Additionally, control experiments where the weight of PDMS molds was monitored after each five-minute wave recording show negligible weight changes, conclusively ruling out any influence of PDMS swelling on our experimental results (Figure~\ref{fig:Swelling}).

\textcolor{black}{We performed numerical simulations of a simplified binary fluid mixture exhibiting Marangoni instabilities (for example, an ethanol–water system) to investigate the evaporation-driven hydrodynamic instability. Convective flows in this system arise from selective evaporation of one component (ethanol) and the resulting surface-tension gradients at the air–liquid interface. In these qualitative simulations—where we assume that the contact line of the fluid mixture is pinned around a single pillar—we show that, in principle, such a simplified ethanol–water mixture can also exhibit a “compositional flower” pattern surrounding the pillar. We stress that these simulations are intended only for qualitative insight and do not quantitatively reproduce the experiments, which involve manyadditional complexities. Moreover, they do not capture thin-film dynamics or fingering instabilities that could emerge once the contact line moves; those phenomena will be addressed in future work.} 

The behavior of a film climbing the exterior of a vertical or an inclined wall, influenced by the opposing forces of a thermally/chemically driven surface tension gradient and gravity, is extensively explored  in the literature through  numerical analysis using a thin-film model (see review articles~\cite{oron1997long,bonn2009wetting,craster2009dynamics} and references therein). In the case of a cylindrical obstacle, these models depend on three parameters: the cylinder radius, the upstream film thickness, and the downstream precursor film thickness. As the  cylinder radius  approaches infinity, the model converges to the scenario of a Marangoni-driven film ascending a vertical plate~\cite{bertozzi1998contact,bertozzi1999undercompressive,munch1999rarefaction,hosoi2001evaporative}. Ref.~\cite{smolka2017dynamics} investigates the stability of the contact line by introducing an infinitesimal two-dimensional perturbation to the traveling wave solution, characterized by sinusoidal variations in the azimuthal direction. Through linear stability analysis, the study determines that the traveling wave (the so called Lax shock) is unstable to long-wavelength perturbations at the contact line. Given the periodic nature of the contact line in the azimuthal direction, only an integer number of fingers can form along it. The analysis identifies the most unstable mode and a neutrally stable cutoff mode, beyond which all modes are stable. These modes increase linearly with the cylinder radius, indicating that the number of fingers formed along the contact line correlates with the cylinder’s circumference. There exists a critical cylinder radius below which the contact line remains stable, with no fingers forming. Our experimental results provide direct validation of these theoretical predictions.  \textcolor{black}{In future work, we will extend the thin-film framework of Smolka et al.\cite{smolka2017dynamics} to rounded-rectangular obstacles—introducing asymmetric evaporation rates and surface-tension gradients—and couple those thin-film equations with the reaction–diffusion equations\cite{field1974oscillations}, to capture the fully coupled fluid–chemical dynamics in the thin film around obstacles of various geometries. }

In our experiments, by tilting the Petri dish, it is possible to selectively control which array of obstacles initially becomes the focal point for wave generation (Figure~\ref{fig:Phase}A-B). As reported in Ref.\cite{agladze1984chaos} and consistent with our experiments, in a tilted Petri dish, waves originate in the thinner areas of the fluid, where convective flows are negligible, and progress toward the thicker regions, where convective flows can develop due to evaporation and the resulting Marangoni forces. The authors of that paper assumed that these convective flows generate a heterogeneous medium where the refractory tail of a preceding wave annihilates part of the following wavefront, resulting in segmented waves that curl inward to form spirals. Over time, a chaotic wave pattern develops. In another study to understand the role of the fluid thickness in an uncovered Petri dish, Rossi et al.\cite{rossi2012segmented} have demonstrated that for fluid layers thinner than 1.0 mm, the BZ system behaves predominantly as a pure reaction-diffusion system, and convective effects are negligible. In layers with an intermediate thickness of 1.0–3.0 mm (the fluid thickness in our experiments was about 1.5 mm), Marangoni instabilities dominate, allowing rippled waves to propagate without breaking, and segmented patterns are not observed. However, for layers thicker than 3.0 mm, buoyancy-driven flows become significant, and wave segmentation processes occur. This resembles the competition between buoyancy-driven flow and solutal Marangoni in evaporating binary droplets \cite{diddens2021competing}. Our experiment using an uncovered, plasma-treated glass Petri dish with a fluid thickness of  2 mm also revealed rippled wavefronts and Marangoni-driven mosaic patterns (Figure~S7A and Video 23)~\cite{miike2010flow,matthiessen1995global,matthiessen1996influence,budroni2009bifurcations}. We notice that at the boundary of the Petri dish, we observed a Marangoni-driven fingering instability, similar to the phenomena we see  near the hydrophilic obstacles. 

Contact angle measurements on both tapered and non-tapered pillars confirm that the contact line remains pinned (Figure~\ref{fig:Partition_New}C-E and~\ref{fig:Cones}C,H), indicating the presence of compensatory upward capillary flows. Additionally, experiments with tapered and inverse-tapered pillars demonstrate that as long as a wetting fluid layer surrounds the obstacle, it continues to function as a wave center. In tapered pillars, we observed that wavefront instability initially appears in straight pillars and, after a time delay, in pillars tapered at approximately 26$^\circ$. This delay is attributed to the higher evaporation rate in straight pillars due to their greater exposed surface area at the contact line.  Interestingly, in a separate experiment investigating the effect of surface roughness, we 3D-printed conical pillars with a wavy surface (Figure~\ref{fig:Roughness}). The results were largely consistent with our standard experiments without surface roughness, except for increased variability in petal size.

Finally, in our experiments with an evaporating CHD-BZ droplet, we observed the formation of spots at the rim. To elucidate the underlying mechanism, identify the more volatile components, and determine which chemical elements are segregated at the boundary, additional comprehensive experiments are necessary. Tracer beads added to the CHD-BZ solution in our evaporating droplet experiment reveal a pinned contact line and the coffee-stain effect, further suggesting the presence of compensatory capillary flows. Interestingly, after approximately one hour, we observed a complex secondary instability at the droplet boundary, which warrants further investigation. Ongoing studies in our laboratory are focused on exploring this phenomenon in more detail.

Our study on the spatio-temporal dynamics of chemical waves interacting with a periodic array of obstacles highlights several potential applications. By demonstrating how these obstacles can synchronize and guide wave propagation, we introduce a strategy for engineering specific wave patterns that could be especially beneficial in advanced microfluidic devices. We further propose that these guided patterns can be harnessed not only for exploring fundamental reaction–diffusion dynamics, but also for practical applications such as chemical computing, where controlled wave interactions serve as the basis for logic operations and information processing \cite{sharma2024programmable}.
Moreover, our findings on manipulating Marangoni-driven flows around hydrophilic obstacles to shape wave patterns can be extended to colloidal droplet evaporation. By employing similar obstacle-based or surface-modification techniques, one could direct evaporative flows and steer colloidal self-assembly, opening new possibilities for creating uniform, tunable deposition patterns in printing, photonics, and related applications \cite{li2024evaporative}. Finally, by adjusting parameters like obstacle geometry or evaporation conditions, one could similarly guide chemical wave propagation within gels, thereby influencing their mechanical oscillations. This capability paves the way for designing advanced biomimetic actuators or “smart” materials where desired wave-driven shape changes (or gel motions) can be precisely programmed or directed~\cite{yoshida2022creation}. Overall, our results provide a foundation for further manipulation of chemical wave behaviors, offering exciting opportunities for technology and manufacturing \cite{yashin2006pattern,kuksenok2007mechanically}.



\section{Method Section}
\label{Method}
\threesubsection{Chemical preparation}
The following chemicals and materials were used in this study. Sodium Bromate (NaBrO$_3$), Sodium Bromide (NaBr), Conc. Sulfuric acid (Conc. H$_2$SO$_4$),  1,4-Cyclohexanedione, Ferrous sulfate, 1,10-phenanthroline was provided by sigma and used without any further purification. Ferroin was synthesized by mixing Ferrous sulfate hexahydrate and 1,10-phenanthroline in 1:3 molar ratio.
For CHD-BZ experiments different concentration of the stock solutions were prepared in dH$_2$O.
Solution A: (494.58 mM solution of sodium bromate)
Solution B: (891.9 mM solution of 1,4-Cyclohexanedione)
Solution C: (971.9 mM solution of NaBr).
A mixture was prepared in a beaker by combining 9 mL of solution A with 1.5 mL of solution B, which was then homogenized using a magnetic stirrer. Following this, 1.5 mL of the Ferroin indicator was added to the mix. Under a chemical fume hood, 0.8 mL of solution C was cautiously introduced to the reaction. The stirring continued for an estimated 1 hour and 45 minutes. Subsequent to this phase, an additional 1.5 mL of Ferroin was incorporated into the mixture. The stirring persisted at a controlled temperature of 23°C for a further 30 minutes, after which it was halted. Attention was then given to the mixture for any signs of color transformation, signaling the start of wave propagation, typically observed 3.5 to 4 hours post the initial mixing process. Following the onset of color change, the CHD-BZ reaction mixture was transferred to a plasma-treated PDMS mold featuring an array of macro-pillars, as specified in Figure~\ref{fig:PetriCloseOpen}A. These pillars, typically 1 mm in diameter and 3 mm high, are arrayed in a square grid with 5 mm spacing. In experiments with variable pillar dimensions and spacings, the volume of the CHD-BZ solution was adjusted to ensure a liquid height of about 1.5 mm, thus avoiding submergence of the pillars. Upon pouring, the solution was swirled gently until the mixture achieved a brown color. The Petri dish was then positioned undisturbed beneath a camera to record the wave propagation, under the illumination of white LED light. For each experiment, we record the waves for 5–10 minutes, observing until the waves from the Petri dish boundary interact with those centered around the obstacles. Afterward, we stop the recording, swirl the solution in the Petri dish to reset the system, and start recording again. Typically, the reagents are fully depleted after about 2 hours. In some experiments, a smaller, modified 50 mm Petri dish was used, and the volume of the CHD-BZ solution was correspondingly reduced to 1.5 mL to maintain the same liquid height. It's important to note that in our standard setups, unless specified otherwise, the pillars do not submerge in the chemical solution. The hydrophilic nature of plasma-treated PDMS causes a thin layer of liquid to ascend the pillars (Figure~\ref{fig:PetriCloseOpen}B), altering the liquid surface near the obstacles as depicted in Figure~\ref{fig:Gold}C. For some control experiments, the surface of the PDMS was coated with a 20 nm titanium layer and then with a pure gold layer. This was done using a Physical Vapor Deposition (PVD) technique with the Lesker Proline PVD 75 apparatus, rotating the substrate four times to achieve uniform coverage on the PDMS pillars' surfaces.
%

\threesubsection{Image Processing}
The wave period is roughly 45 sec, and we capture images of the chemical waves at a rate of 1 frame per second across multiple wave cycles. To minimize spatial noise, we average the captured images over several wave cycles, typically 6-8, and then subtract each individual image from the time-averaged image. After this, we apply a Gaussian blur filter and use a customized MATLAB code to perform the Hilbert transform, which helps us calculate the phase map. This code is provided in the Supplementary Information.

{\color{black}{
\threesubsection{Numerical Simulations}
Our numerical simulations qualitatively capture the hydrodynamic instabilities observed in CHD-BZ experiments. Our aim is to demonstrate that solutal Marangoni–driven instabilities, induced by evaporation, suffice to explain the symmetry-breaking patterns seen experimentally. We do not attempt a quantitative match to the data, as that would require a substantially more complex model.

We model the system using the experimental geometry: a binary mixture exhibiting Marangoni instabilities -- one example being an ethanol–water system -- is confined around a cylindrical pillar of radius \(R = \SI{1}{\milli\metre}\), height of \(\SI{5}{\milli\metre}\), and center-to-center spacing of \(\SI{7}{\milli\metre}\) (see Figure \ref{fig:geometry}). At the evaluation time \(t = 599\)\,s—recorded from our experiments—the liquid layer meets the pillar walls at a contact angle \(\theta = 52^\circ\) and has a height of \(\SI{2.5}{\milli\metre}\) there. The evaporation is assumed to be limited by vapor diffusion, a reasonable assumption for liquids sufficiently below the boiling point. The contact line remains pinned to the pillar walls, and a symmetry boundary is imposed at the midpoint between adjacent pillars. Evaporation is modeled using the quasi-stationary framework of Rocha et al.~\cite{rocha2024marangoni}, suitably adapted to this geometry.

We track the mass fraction \(y_A\) of the more volatile component \(A\) (e.g., ethanol), with \(y_B = 1 - y_A\) denoting the mass fraction of the less volatile component \(B\) (e.g., water). For slowly evaporating liquids, compositional variations remain small, so we decompose
\begin{equation}
y_A(x,t) = y_{A,0}(t_0) + y(x,t)\,,
\label{eq:qsmassfractiondecomposition}
\end{equation}
where \(y_{A,0}(t_0)\) is the spatially averaged mass fraction at the evaluation time \(t_0\) and \(y(x,t)\) captures local deviations.

Although \(y_{A,0}(t)\) evolves over time, the flow can be treated as quasi‐steady at each drying stage, provided the evaporation rate remains low. All material properties are held constant except for surface tension, which is a function of the mass fraction variations. We assume a linear dependence
\[
\sigma(y_A) = \sigma\bigl(y_{A,0}(t)\bigr) + y\,\frac{\partial\sigma}{\partial y_A}\,.
\]

The system is nondimensionalized as follows:
\[
x = R\,\tilde x,\quad
t = \frac{R^2}{D_0}\,\tilde t,\quad
\mathbf u = \frac{D_0}{R}\,\tilde{\mathbf u},\quad
p = \frac{\mu_0\,D_0}{R^2}\,\tilde p,\quad
c_\alpha - c_\alpha^\infty = \bigl(c_\alpha^\text{eq} - c_\alpha^\infty\bigr)\,\tilde c_\alpha,
\]
where \(D_0\) and \(\mu_0\) are the diffusion coefficient and viscosity at the mean composition,  
\[
c_\alpha^\text{eq} = c_\alpha^\text{pure}\,\gamma_\alpha\,x_\alpha
\]
is the equilibrium vapor concentration (with \(c_\alpha^\text{pure}\) the pure‐component saturation, \(\gamma_\alpha\) the activity coefficient, and \(x_\alpha\) the mole fraction), and \(c_\alpha^\infty\) is the far‐field vapor concentration.

Assuming small local variations in \(y\), we treat \(c_\alpha^\text{eq}\) as constant \cite{diddens2021competing}. Under this scaling, the dimensionless vapor field \(\tilde c\) in the gas phase satisfies the Laplace equation
\begin{equation}
	\tilde\nabla^2 \tilde c = 0
\end{equation}
subject to the boundary conditions \(\tilde c = 0\) at the liquid–vapor interface and \(\tilde c = 1\) in the far field \cite{deegan1997capillary,popov2005evaporative,diddens2021competing}. The solution \(\tilde c\) then maps directly back to the physical vapor concentration \(c_\alpha\). The dimensionless evaporation rate $\tilde{j}$ is given by the diffusive mass flux
\begin{equation}
	\tilde{j} = -  \mathbf{n} \boldsymbol{\cdot} \tilde{\boldsymbol{\nabla}} \tilde{c} ,
\end{equation}
with $\mathbf{n}$ being the outward normal vector at the liquid-vapour interface.
Evaporation drives variations in the liquid composition through the flux boundary condition:
\begin{equation}\label{eq:evaporation_bc}
	- \tilde{\boldsymbol{\nabla}} y \boldsymbol{\cdot} \mathbf{n} = \mathrm{Ev} \tilde{j},
\end{equation}
where $\mathrm{Ev}$ is the evaporation number, which was originally defined by \cite{diddens2021competing} as
\begin{equation}
	\mathrm{Ev} =  \dfrac{(1 - y_{A,0})D_A^\mathrm{vap}(c_{A,0}^\mathrm{eq} - c_A^\infty)-y_{A,0} D_B^\mathrm{vap}(c_{B,0}^\mathrm{eq} - c_B^\infty)}{\rho_0 D_0},
\end{equation}
where $D_A^\mathrm{vap}$ and $D_B^\mathrm{vap}$ are the vapour diffusion coefficients of components $A$ and $B$, respectively, and $\rho_0$ is the liquid density evaluated at the spatially averaged composition.

In the liquid phase, we consider the incompressible Stokes equations (neglecting inertia and gravity):
\begin{equation}
	\tilde{\boldsymbol{\nabla}} \boldsymbol{\cdot} \tilde{\mathbf{u}} = 0,
\end{equation}
\begin{equation}
	\tilde{\boldsymbol{\nabla}} p = \tilde{\boldsymbol{\nabla}}^2 \tilde{\mathbf{u}}.
\end{equation}
In order to reduce the number of parameters in the problem, we define $\xi = y / \mathrm{Ev}$, such that Eq.~\eqref{eq:evaporation_bc} becomes:
\begin{equation}\label{eq:evaporation_bc2}
	- \tilde{\boldsymbol{\nabla}} \xi \boldsymbol{\cdot} \mathbf{n} = \tilde{j}.
\end{equation}
The evolution of $\xi$ in the liquid phase is then given by the advection-diffusion equation:
\begin{equation}
	\partial_{\tilde{t}} \xi + \tilde{\mathbf{u}} \boldsymbol{\cdot} \tilde{\boldsymbol{\nabla}} \xi = \tilde{\nabla}^2 \xi.
\end{equation}
Marangoni stresses are considered at the liquid-vapour interface via the boundary condition:
\begin{equation}\label{eq:marangoni_stress}
	\mathbf{n} \boldsymbol{\cdot} (\tilde{\boldsymbol{\nabla}} \tilde{\mathbf{u}} + (\tilde{\boldsymbol{\nabla}}\tilde{\mathbf{u}})^t) \boldsymbol{\cdot} \mathbf{t} = \mathrm{Ma} \tilde{\boldsymbol{\nabla}}_t \xi \boldsymbol{\cdot} \mathbf{t},
\end{equation}
with $\mathbf{t}$ being the tangential vector at the liquid-vapour interface and $\mathrm{Ma}$ being the solutal Marangoni number, defined as
\begin{equation}
	\mathrm{Ma} = \frac{R \partial_{y_A} \sigma}{\mu_0 D_0} \mathrm{Ev}.
\end{equation}

We consider that the shape of the liquid layer does not change due to surface tension effects (i.e.\ capillary number $\mathrm{Ca} \rightarrow 0$).
At the liquid-gas interface, we consider that evaporation is slow, i.e.\ $\mathbf{u}_I \ll \mathbf{u}$, where $\mathbf{u}_I$ is the velocity at the interface and $\mathbf{u}$ is the velocity in the bulk liquid.
This allows us to assume that the interface is stationary, i.e.\ we impose the boundary condition $\tilde{\mathbf{u}} \boldsymbol{\cdot} \mathbf{n} = 0$ at the liquid-vapour interface, which we enforce via a locally adjusted normal traction.
The volume of liquid in the system is fixed to a value such that the contact angle is $\theta = 52^\circ$ at the pillar walls and the liquid is pinned to the pillar walls.
At the substrate, we impose a no-slip boundary condition $\tilde{\mathbf{u}} = 0$ and a no-penetration boundary condition $\boldsymbol{\nabla} \xi \boldsymbol{\cdot} \mathbf{n} = 0$.
We constraint the spatial average of $\xi$ to zero, in accordance with the decomposition \eqref{eq:qsmassfractiondecomposition}.

The system of equations is solved using the finite element method with the open-source library \texttt{pyoomph} \cite{diddens2024bifurcation}, which is based on \texttt{oomph-lib} \cite{heil2006oomph} and \texttt{GiNaC} \cite{bauer2002introduction}.
All domains are discretised using triangular elements. 
The operators in the system of equations are given in a cylindrical coordinate system.
Linear basis functions are used for $\xi$ and $\tilde{p}$, whilst quadratic basis functions are used for the velocity field $\tilde{\mathbf{u}}$.
We follow the methodology outlined in \cite{rocha2024marangoni} to compute the steady base state of the system and its stability against perturbations characterised by different azimuthal wavenumbers $m$.
This allows us to assess whether the system's symmetry is broken and forms patterns with a specific azimuthal wavenumber, like those observed in the experiments.

The  stability of the simplified system depends on the Marangoni number $\mathrm{Ma}$, on the distance between pillars $D$, on the angle $\theta$ at the pillar walls, and on the liquid height at the pillar walls.
In the following, for simplicity, we fix the liquid height at $2.5$ mm, $D = 7$ mm, $\theta = 52^\circ$, and vary $\mathrm{Ma}$.

}}
\newpage
\medskip
\noindent\textbf{Supporting Information} \par 
\noindent Supporting Information is available from the Wiley Online Library or from the author.

\medskip
\noindent\textbf{Acknowledgements} \par 
\noindent We deeply appreciate the valuable insights and discussions provided by Professors G. Shubeita, A. Narayanan, A. Rebane, A. Pumir, O. Omelchenko and E. Frey. We also thank  Dr. I. Gholami and M. Ibrahim  for technical assistance and to the anonymous reviewers for their thoughtful comments and suggestions. We are especially grateful to the Core Technology Platform (CTP) at NYUAD for their exceptional support, with special thanks to Dr. R. Rezgui for his critical assistance with microscopy, Dr. Q. Zhang for his support in the clean room, and V. Dhanvi and J. Govindan for their outstanding contributions to fabricating the acrylic molds. We also acknowledge the High Performance Computing (HPC) facility at NYUAD for providing essential resources for data storage and analysis, which were integral to this study. Finally, this work was also supported by an Industrial Partnership Programme, High Tech Systems and Materials (HTSM), of the Netherlands Organisation for Scientific Research (NWO); a funding for public-private partnerships (PPS) of the Netherlands Enterprise Agency (RVO) and the Ministry of Economic Affairs (EZ); Canon Production Printing Netherlands B.V.; University of Twente; and Eindhoven University of Technology (project TKI HTSM - CANON - P1 - PRINTHEAD \& DROPLET FORMATION, grant no. PPS2107). \\
\noindent\textbf{Author contributions} \par 
\noindent S.G. and B.P. conducted the experiments, D.R. , C.D. and D.L. designed and carried out the numerical simulations. A.G. conceived the idea, designed the experiments, analyzed the experimental data, and wrote the first draft of the manuscript. All authors discussed the results and revised the manuscript.\\
\noindent\textbf{Data and materials availability} \par
\noindent All data needed to evaluate the conclusions in the paper are present in the paper and/or the Supplementary Materials. Additional data related to this paper may be requested from the authors.

\medskip

%
\bibliographystyle{MSP}
\bibliography{bibliography}

\newpage
\renewcommand{\thepage}{S\arabic{page}}
\renewcommand{\thesection}{S\arabic{section}}
\renewcommand{\thetable}{S\arabic{table}}
\renewcommand{\thefigure}{S\arabic{figure}}
\setcounter{equation}{0} 
\setcounter{page}{0} 
\renewcommand{\theequation}{S.\arabic{equation}}
  \setcounter{figure}{0}
\def\fnum@figure{\figurename\thefigure}
\newpage
\section*{Supplemental Figures}
\begin{figure}[htbp!]
\begin{center}
	\includegraphics[width=0.7\columnwidth]{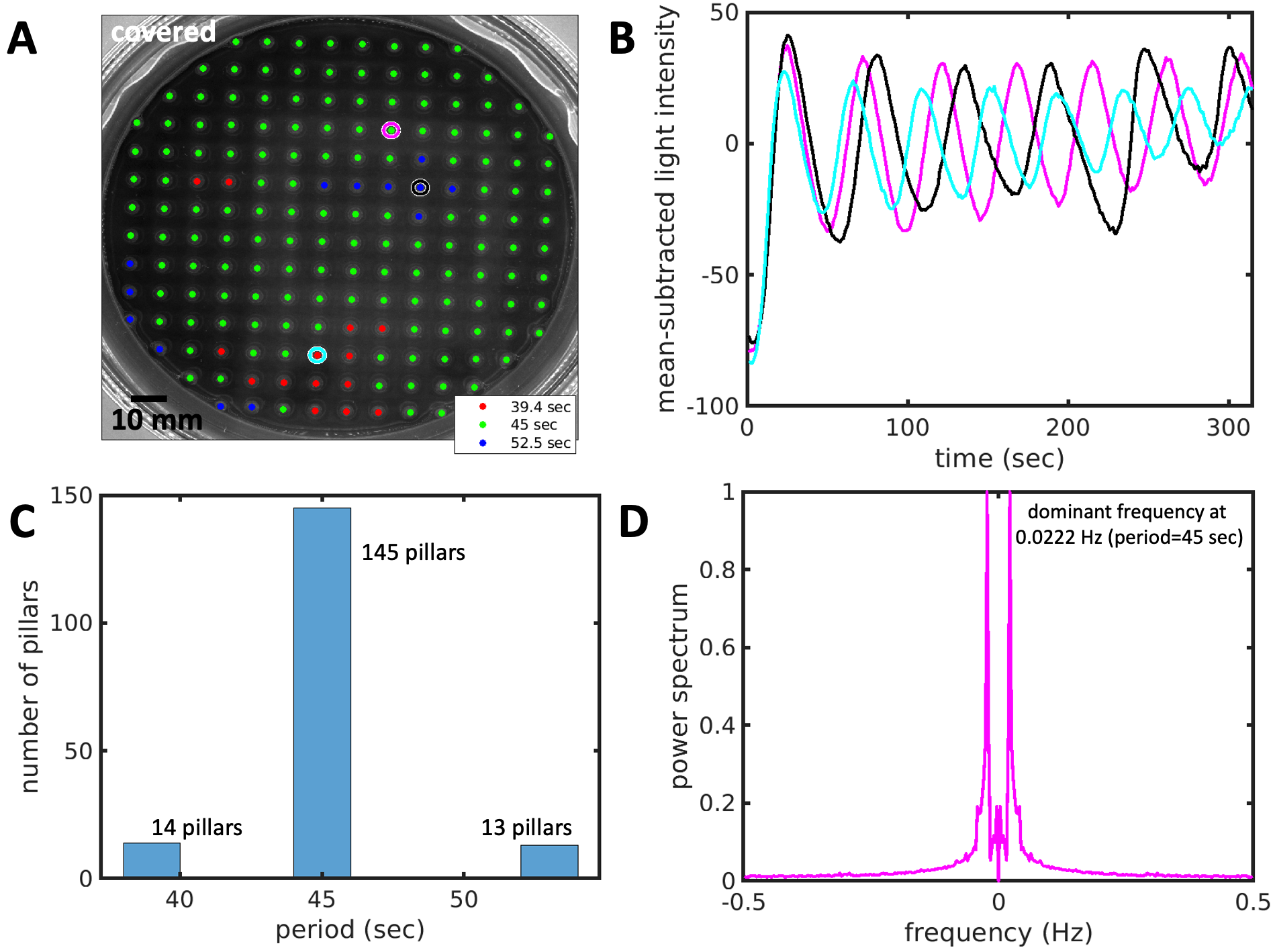}
	\includegraphics[width=0.7\columnwidth]{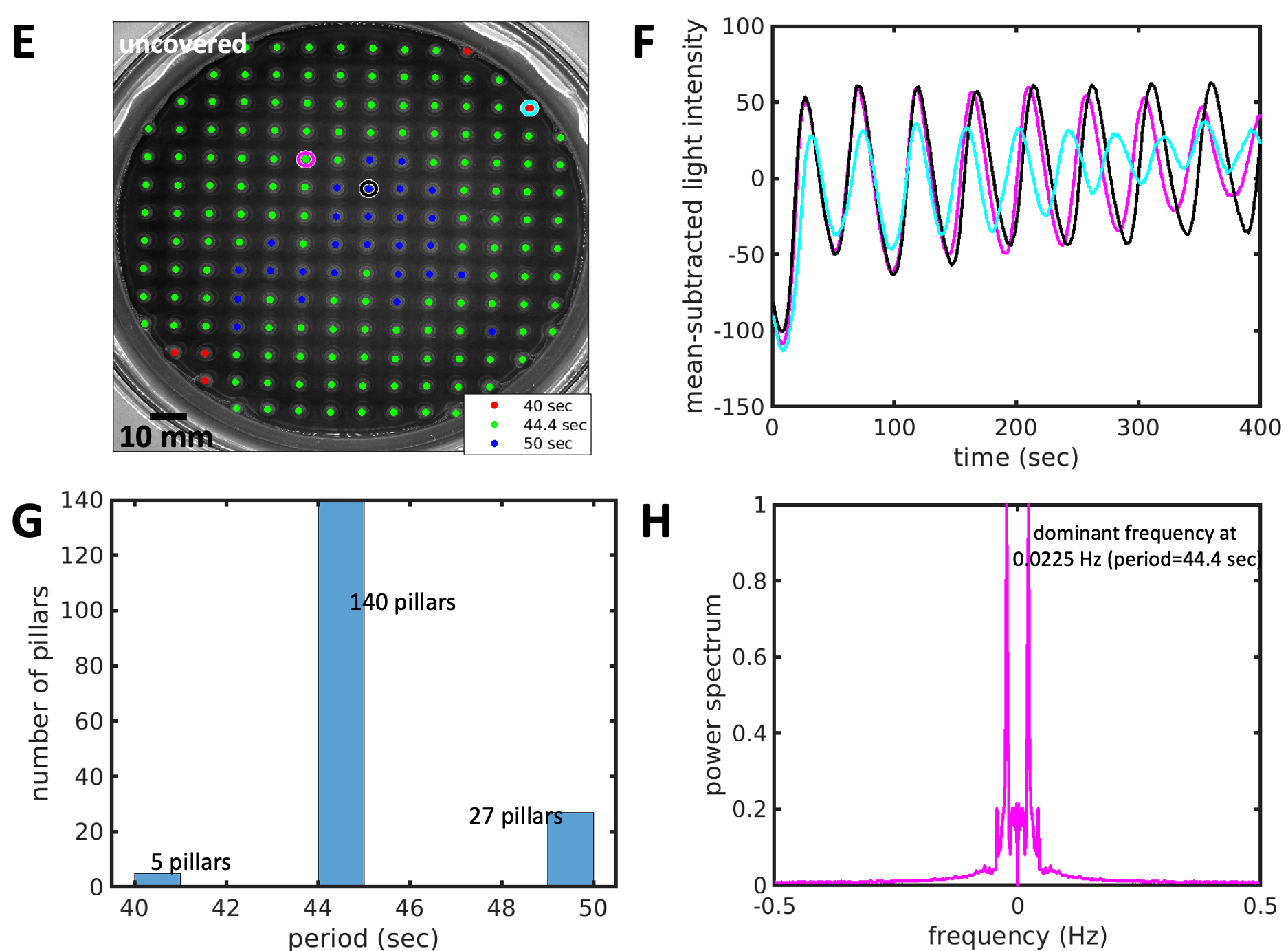}
	\caption{\textbf{Frequency Analysis}. (\textbf{A}) Color-coded period of the waves centered around the obstacles, measured in seconds. The Petri dish is covered and the waves maintain their circular shapes over multiple cycles, as shown in Figure~\ref{fig:PetriCloseOpen}K. (\textbf{B}) Averaged light intensity oscillations around three selected pillars highlighted in panel A, showing oscillations with periods of 45 seconds (magenta), 52.5 seconds (black), and 39.4 seconds (cyan).  (\textbf{C}) Histogram of oscillations period. (\textbf{D}) Power spectrum of the oscillations shown in magenta in panel (B). (\textbf{E}-\textbf{H}) Similar to panels (A-D), but with the Petri dish uncovered, affecting the wave dynamics as shown in Figure~\ref{fig:PetriCloseOpen}L. }
	\label{fig:frequency}
\end{center}
\end{figure}
\begin{figure}[htbp!]
\begin{center}
	\includegraphics[width=0.8\columnwidth]{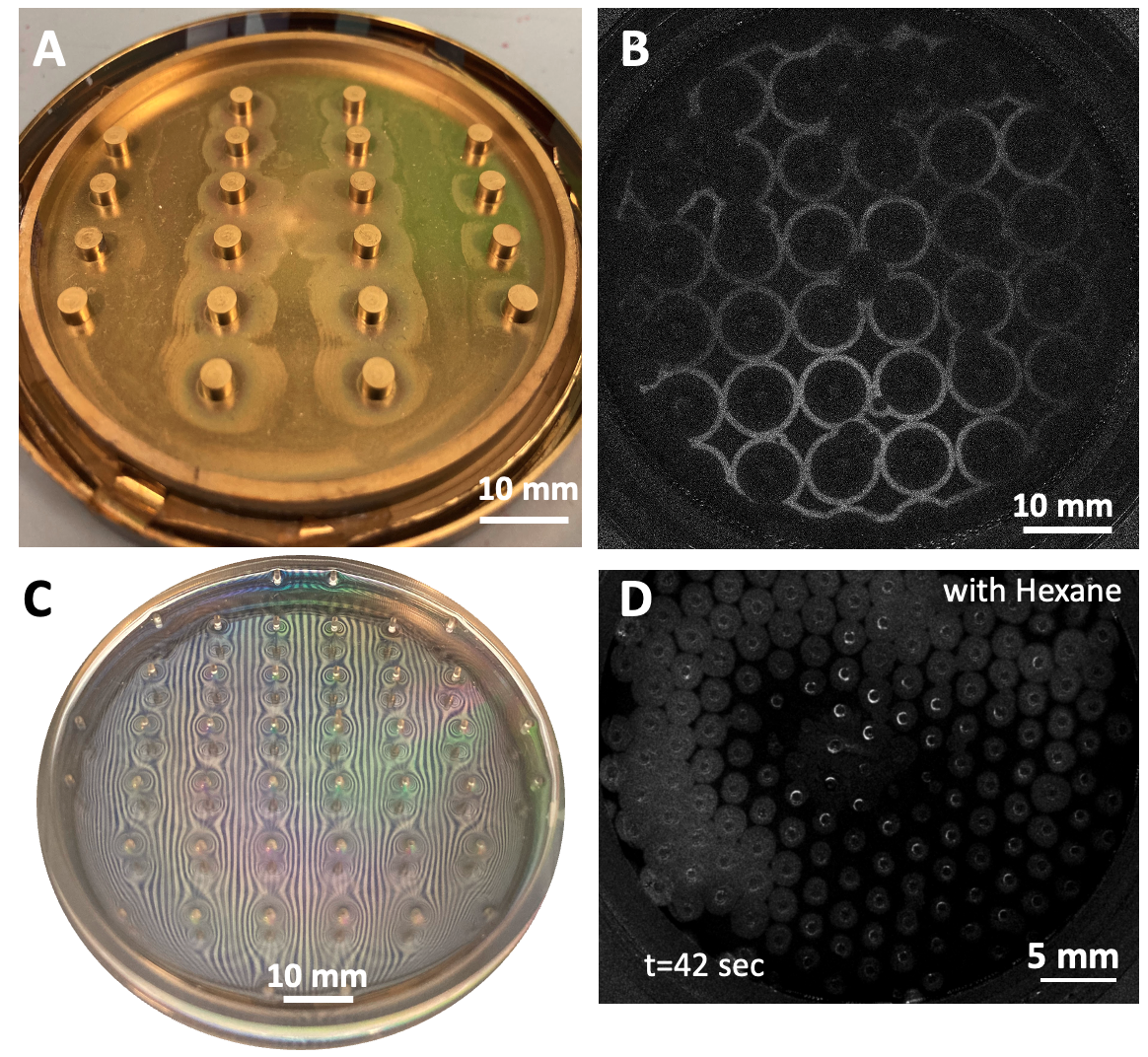}		
	\includegraphics[width=0.5\columnwidth]{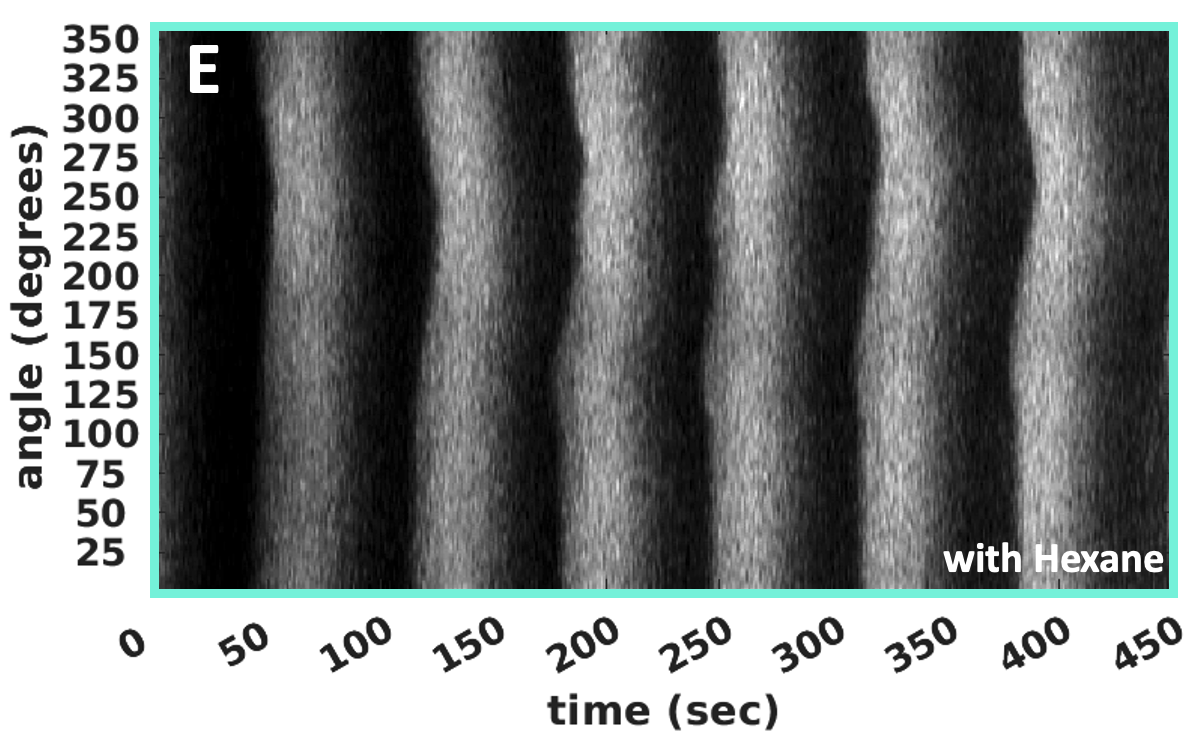}		
	\caption{ (\textbf{A}) A sample of a PDMS mold coated with a thin layer of gold is shown. The large pillars have a diameter of 3.8 mm. (\textbf{B}) Synchronous waves are centered around the gold-coated pillars. (\textbf{C}) A striped pattern is projected onto the fluid surface, highlighting the deformation caused by the rising fluid around the hydrophilic pillars. \textbf{(D)} A thin layer of hexane (with about 1 mm thickness) over the CHD-BZ solution prevents evaporation, ensuring the wavefronts preserve their circular shapes. Please note that the obstacles are not submerged. \textbf{(E)} Space-time plot along the cyan circle in panel D illustrating that over time the wavefronts maintain their circular shape. }
	\label{fig:Gold}
\end{center}
\end{figure}
\begin{figure}[htbp!]
\begin{center}
	\includegraphics[width=\columnwidth]{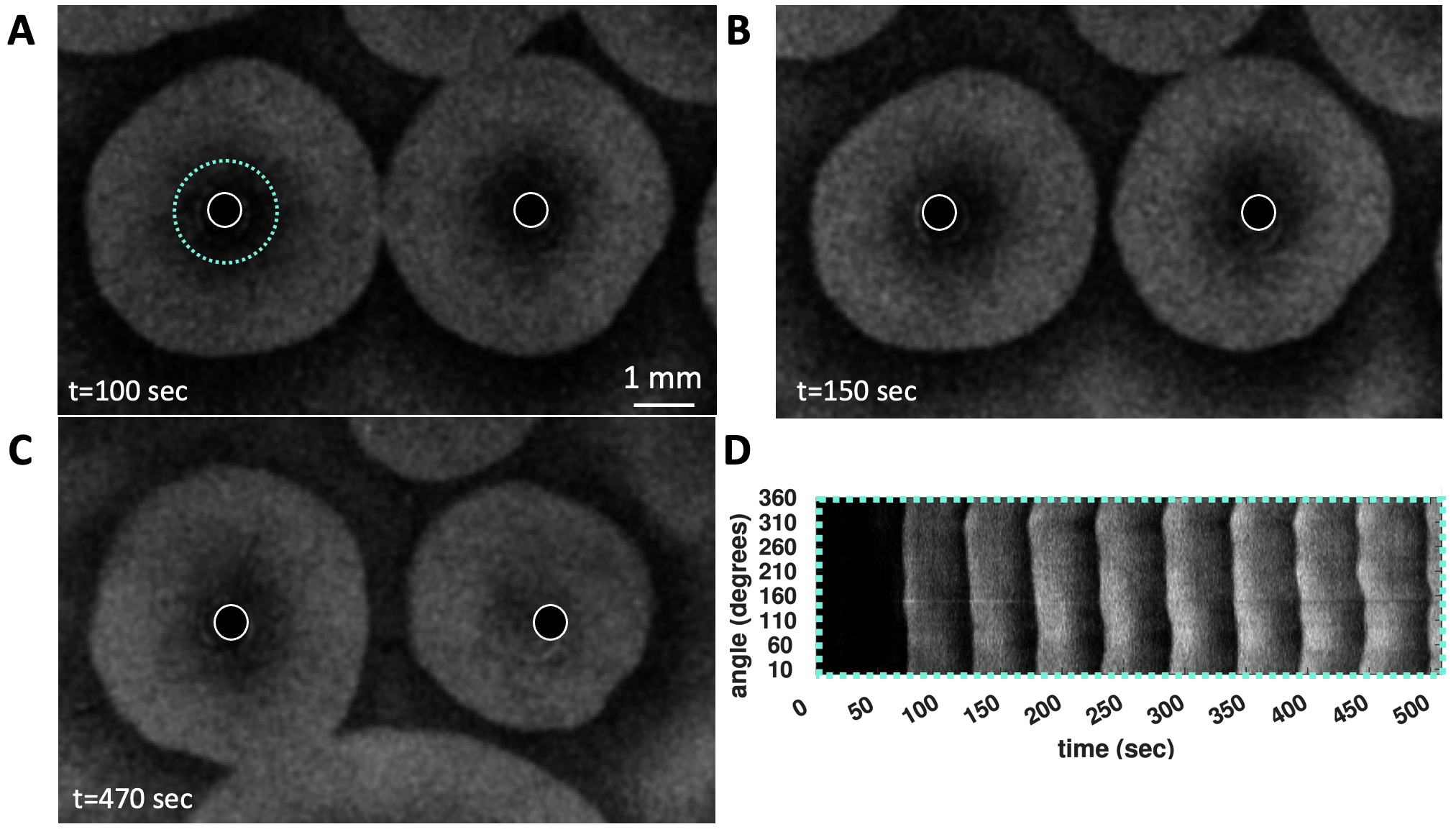}
	\caption{\textbf{No fingering instability in small-diameter pillars}. (\textbf{A}-\textbf{D}) In experiments conducted with uncovered Petri dishes featuring pillars of 0.5 mm diameter, there is no occurrence of wavefront instability. This observation indicates that a certain minimum perimeter of the pillars is necessary for fingering instability to occur. The space-time plot shown in panel (D) is created by accumulating the light intensity around the cyan-colored circle depicted in panel (A).}
	\label{fig:ThinPillars}
\end{center}
\end{figure}
\begin{figure}[htbp!]
\begin{center}
	\includegraphics[width=0.9\columnwidth]{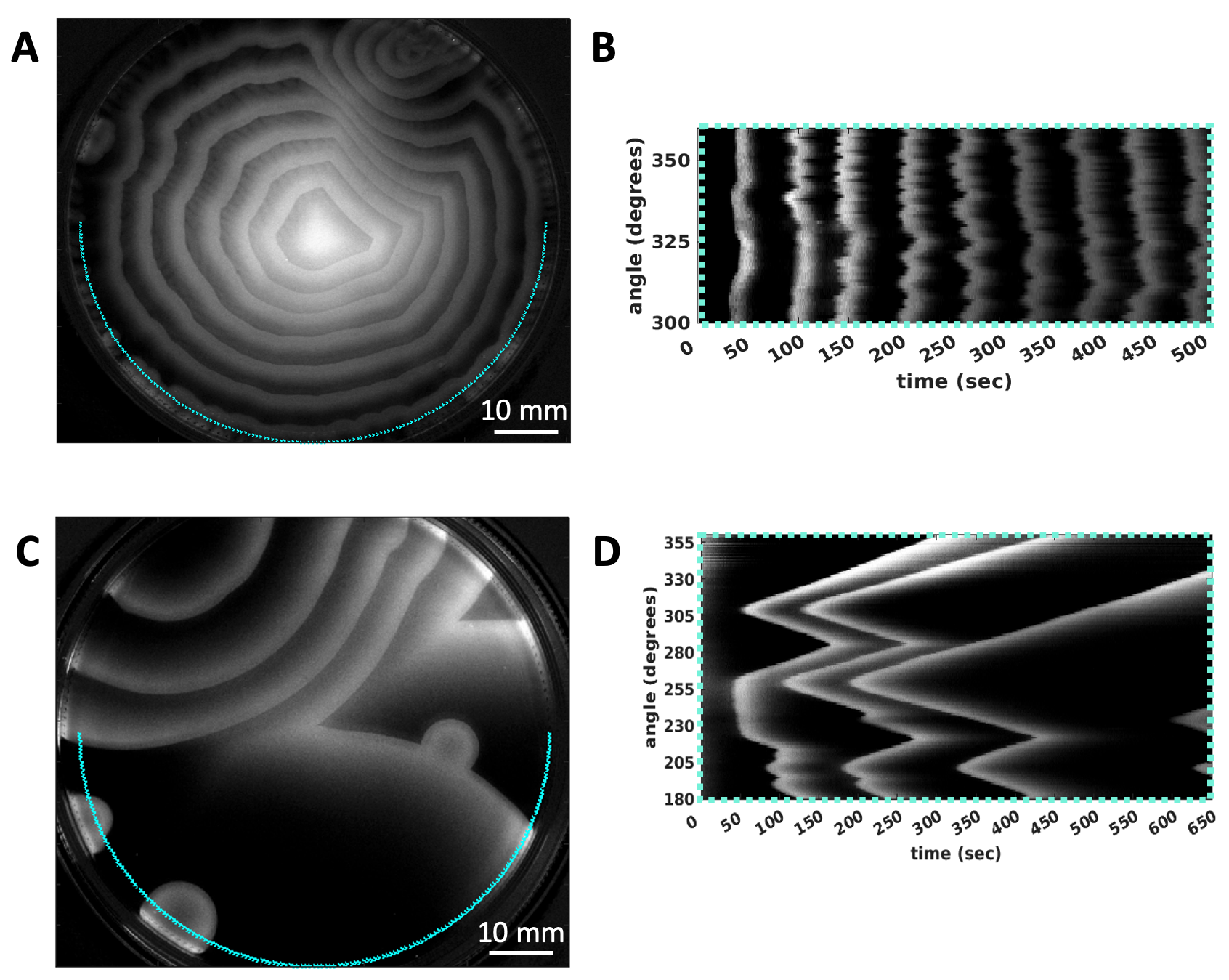}
	\caption{\textbf{Experiment in a glass Petri dish without obstacles}. (\textbf{A}) A top view of a glass uncovered Petri dish with plasma treatment, illustrating fingering instability at the glass boundary. Notice that rippled waves travel and make visible the underlying mosaic patterns generated by the Marangoni instabilities. (\textbf{B}) A space-time plot along the cyan-colored semicircle shows the dynamics of the fingering, emphasizing that the positions of the fingers can change over time. (\textbf{C-D}) A setup similar to panel (A), but covered to minimize evaporation. In this configuration, wave centers appear at the periphery, but the wavefronts maintain their circular forms. }
	\label{fig:Glass}
\end{center}
\end{figure}
\begin{figure}[htbp!]
\begin{center}
	\includegraphics[width=0.8\columnwidth]{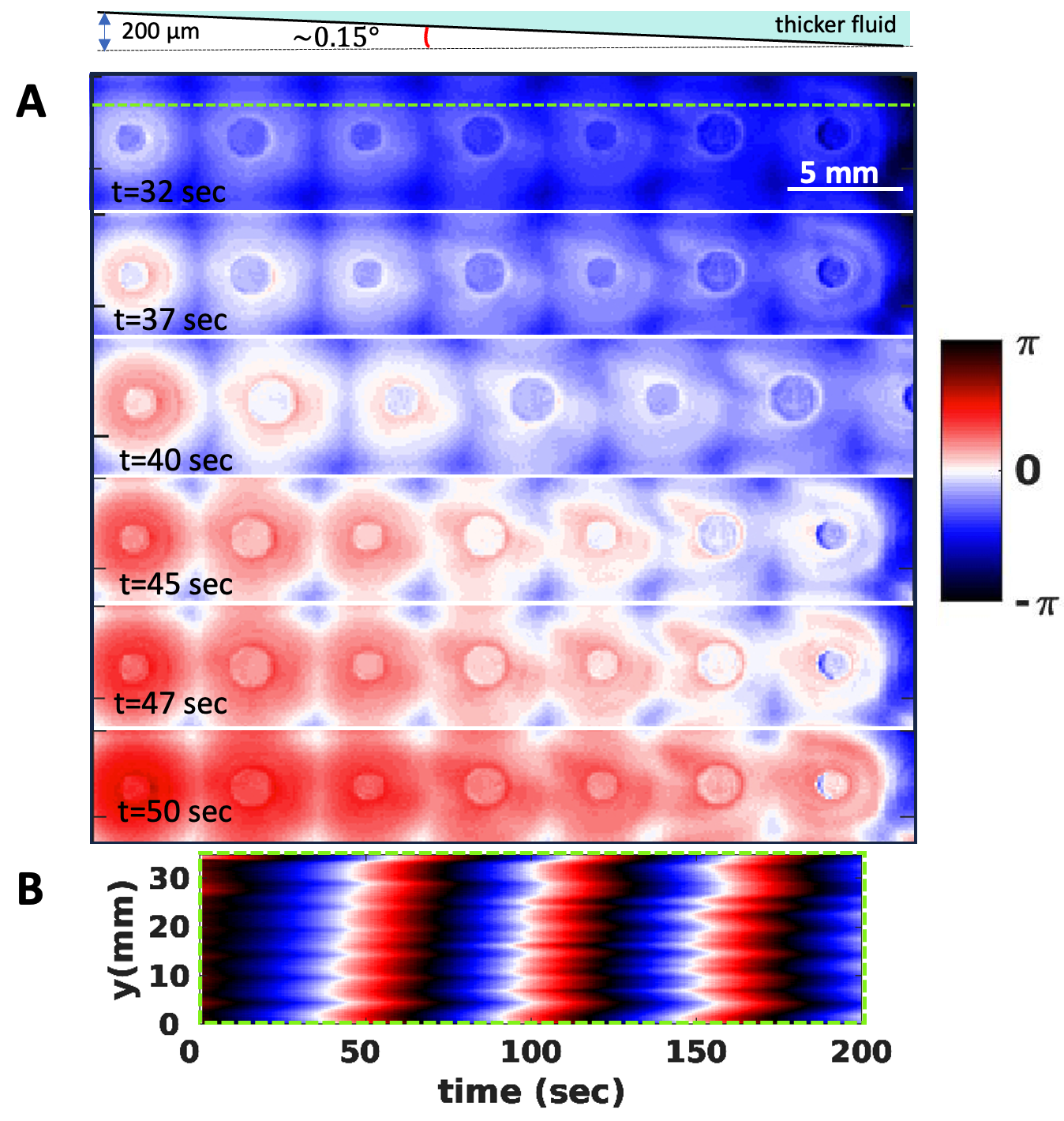}
	\caption{\textbf{Phase map of an experiment in a tilted Petri dish} (\textbf{A}) In a Petri dish with a slight tilt, synchronization waves initiate in the thinner part of the fluid and propagate towards the area with greater fluid thickness (video 15). (\textbf{B}) Space-time plot of the phase map along the green dashed line indicated in panel (A) illustrates the propagation of phase waves. Notice that in this experiment, the Petri dish is not covered, and flower patterns emerge at $t\sim80$ sec.}
	\label{fig:Phase}
\end{center}
\end{figure}
\begin{figure}[hbpt!]
\begin{center}
	\includegraphics[width=\columnwidth]{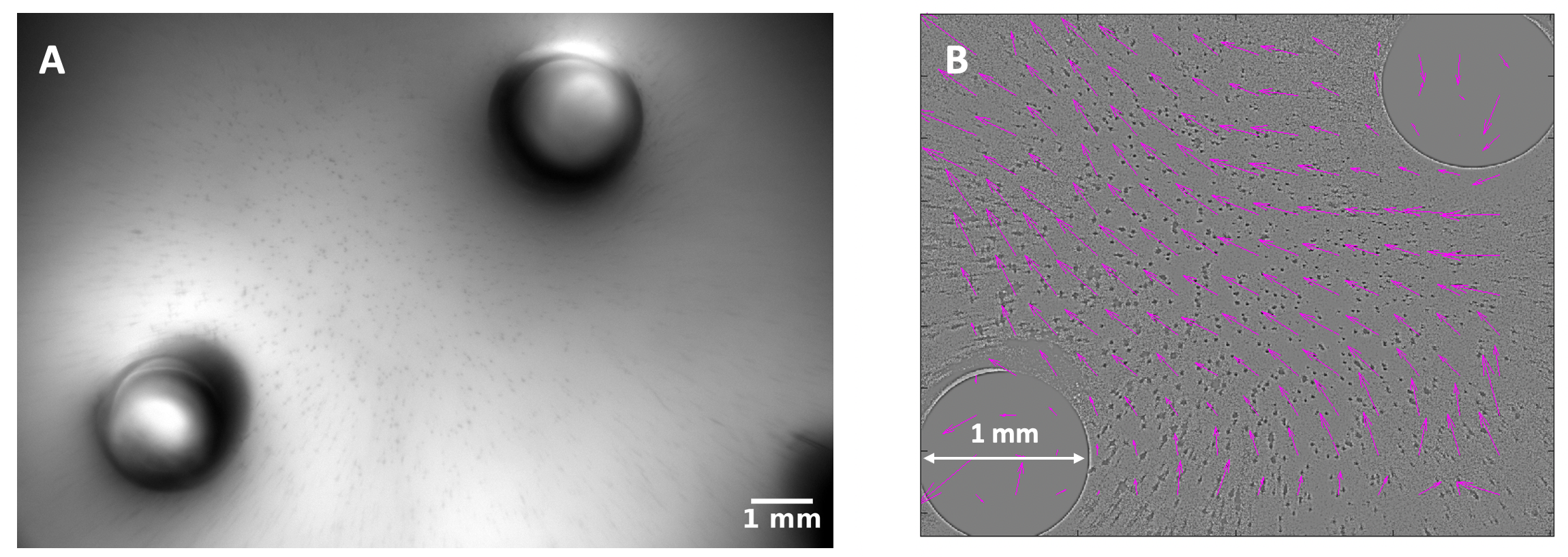}
	\caption{\textbf{Marangoni-driven surface flows.} \textbf{(A-B}) In this experiment, tracer particles (20 $\mu$m in diameter) are used to visualize surface flows driven by Marangoni flows. These flows arise from chemical concentration gradients and slight temperature differences caused by evaporative cooling, which generate surface tension gradients (refer to video 27). }
	\label{fig:BeadTracking}
\end{center}
\end{figure}
\begin{figure}[htbp!]
\begin{center}
	\includegraphics[width=0.8\columnwidth]{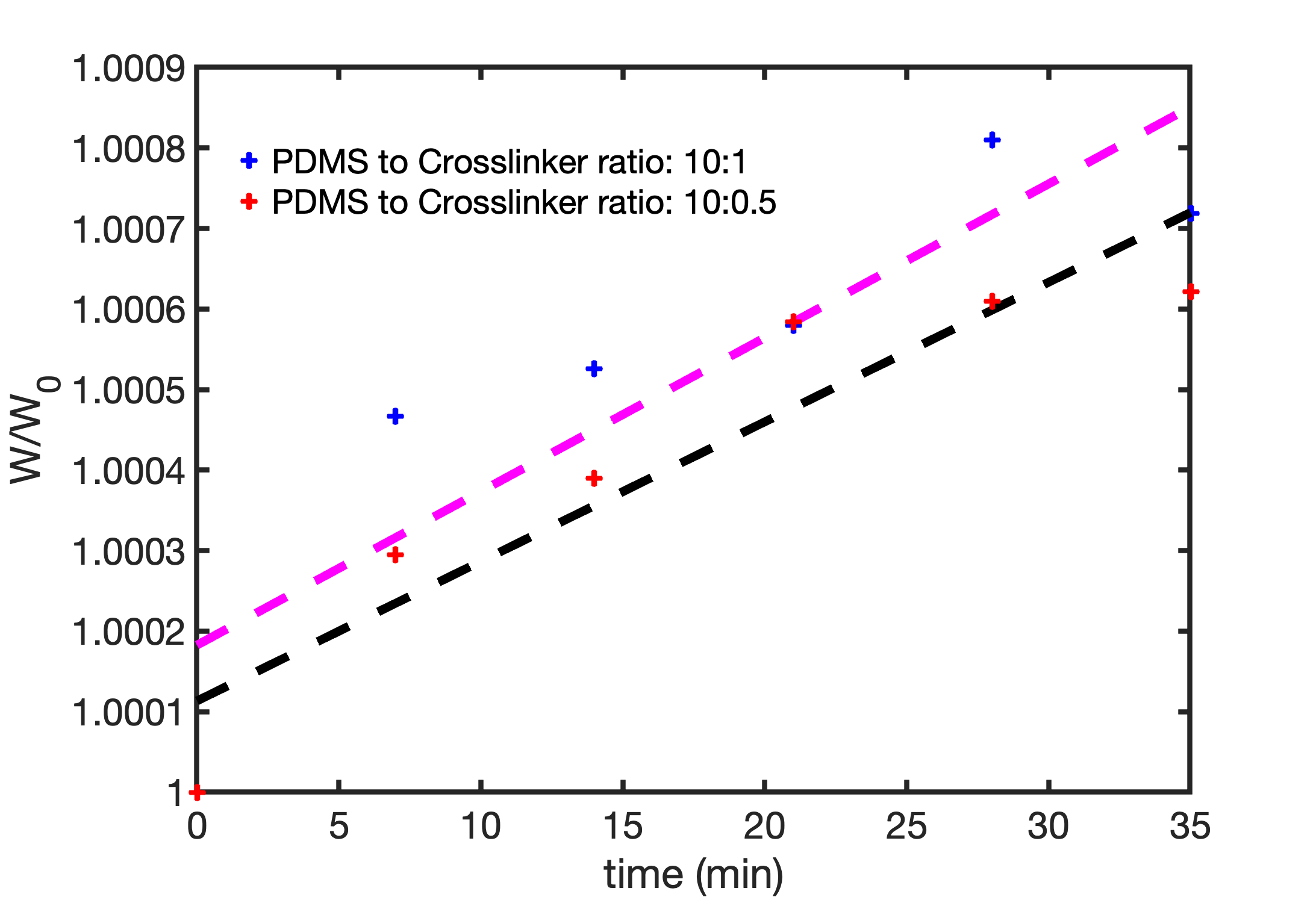}
	\caption{\textcolor{black}{\textbf{Negligible swelling of PDMS:} To rule out PDMS swelling during our experiments, we measured the weight of the PDMS after each recording and found negligible changes. The initial PDMS weight at time zero is denoted as $W_0$. Data are presented for different PDMS-to-cross-linker ratios. All experiments were conducted using a PDMS-to-cross-linker ratio of 10:1. Even with a softer PDMS ratio of 10:0.5, the swelling effect remained negligible.}}
	\label{fig:Swelling}
\end{center}
\end{figure}
\begin{figure}[htbp!]
\begin{center}
	\includegraphics[width=\columnwidth]{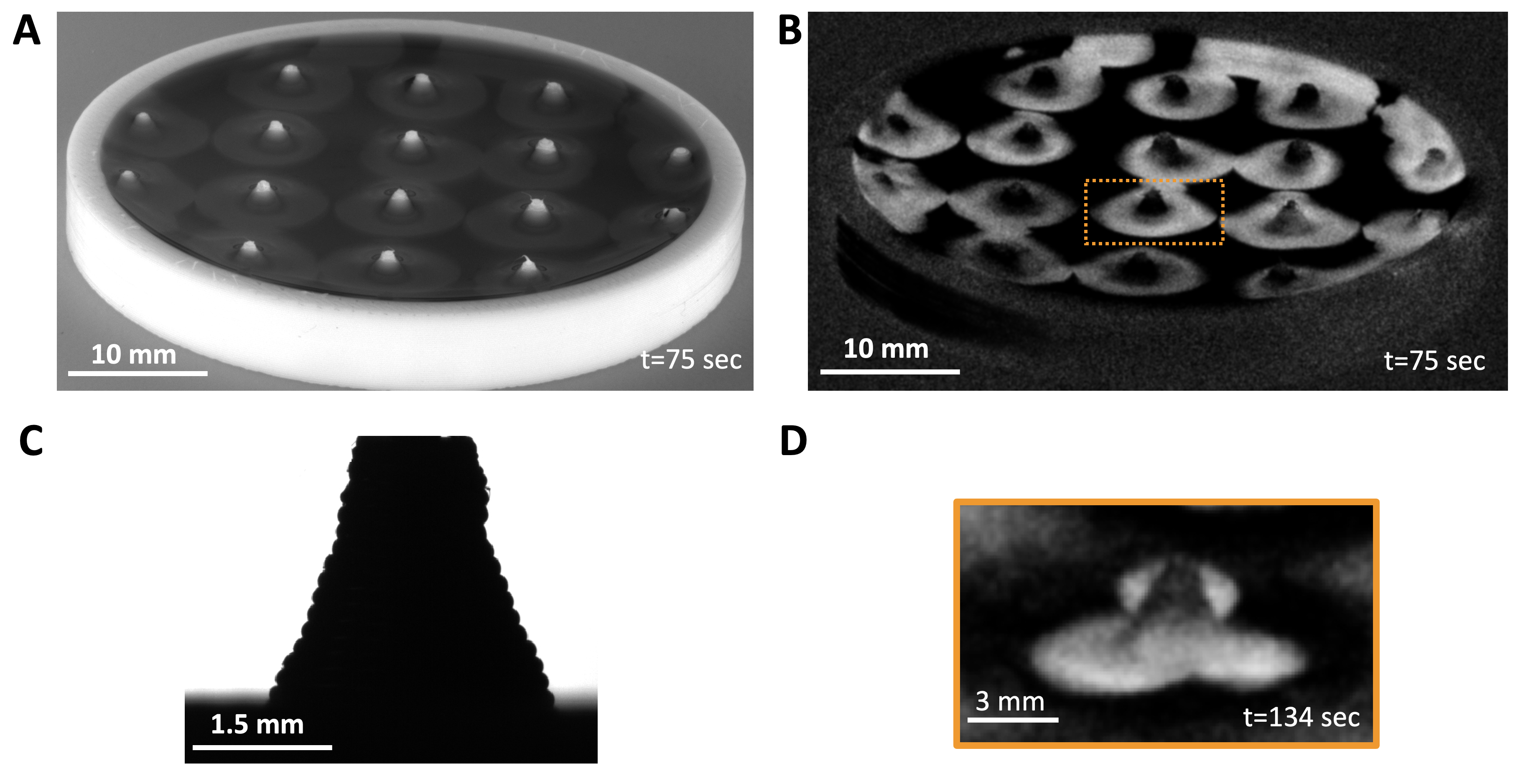}
	\caption{\textcolor{black}{\textbf{Experiments with rough obstacles:} To examine the impact of surface roughness, we used a conical array of 3D-printed PLA (polylactic acid) obstacles with a wavy surface. Despite their rough texture, these obstacles continued to act as wave centers. In an open setup, the circular wavefronts fragmented, forming flower-like patterns, though with greater variability in petal size.	The image in panel B is a processed version of the image in panel A, where the temporal mean has been subtracted, and a Gaussian blur has been applied.	}}
	\label{fig:Roughness}
\end{center}
\end{figure}

\newpage
\color{black}{

\section*{Meniscus analysis}
At static equilibrium, the pressure jump across a liquid--air interface equals surface tension times mean curvature (Young--Laplace law):
\begin{equation}
	\gamma\,\kappa \;=\; \Delta P.
	\label{eq:YL}
\end{equation}
Measuring height $H(r)$ upward (relative to a far-field reference) gives the hydrostatic pressure difference $\Delta p=\rho g\,H$, where $\rho$ is the liquid--air density difference and $g$ is gravity. For an axisymmetric meniscus $z=H(r)$ around a vertical cylinder, the exact mean curvature is the sum of principal curvatures:
\begin{equation}
	\kappa \;=\; \,\frac{H''}{\bigl(1+H'^2\bigr)^{3/2}} \;+\; \frac{1}{r}\,\frac{H'}{\sqrt{1+H'^2}},
	\label{eq:kappa_exact}
\end{equation}
where primes denote derivatives with respect to $r$.

\begin{figure}[t!]
	\begin{center}
		\includegraphics[width=\columnwidth]{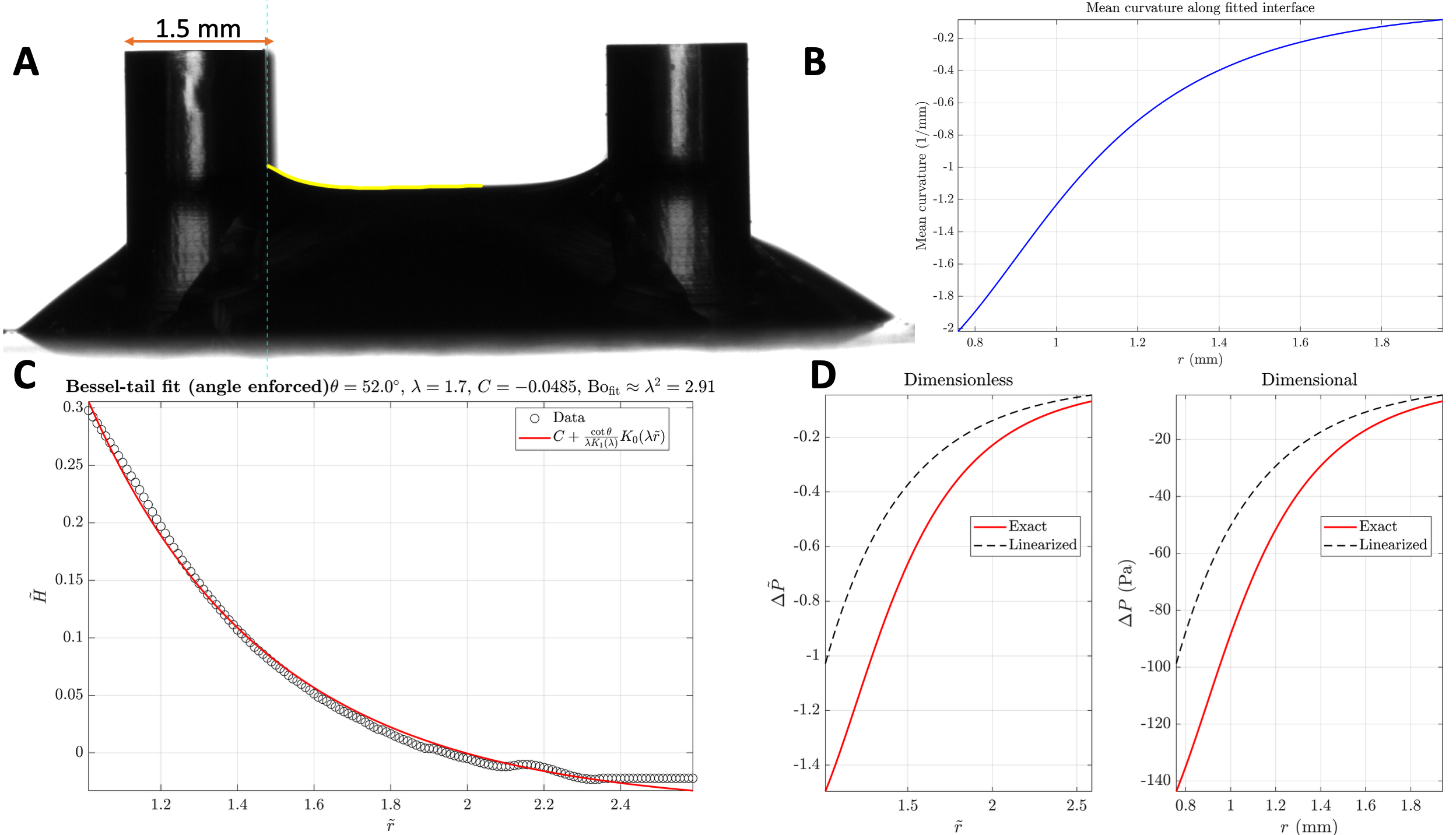}
		\caption{\color{black}{\textbf{Meniscus extraction, model fit, and pressure reconstruction.} (A) Side-view of the rising film around a cylindrical obstacle; the interface (yellow) is obtained by edge tracing within a user-defined region of interest, and the dashed line marks the cylinder wall. (B) Dimensionless meniscus profile (circles) together with the best-fit far-field model that enforces the wall contact angle; the example fit returns lambda 1.7, offset -0.0485, and a fitted Bond number of 2.91. (C) Laplace pressure reconstructed along the interface, shown in both dimensionless and dimensional units; the solid red curve uses the exact curvature while the dashed black curve shows a linearized approximation, with the largest differences near the wall.}}
		\label{fig:meniscus}
	\end{center}
\end{figure}

Let $R$ be the cylinder radius and define nondimensional variables $\tilde r = r/R$ and $\tilde H = H/R$. Combining~\eqref{eq:YL}--\eqref{eq:kappa_exact} yields the exact (nonlinear) shape equation
\begin{equation}
	\frac{d}{d\tilde r}\left[\frac{\tilde r\,\tilde H'}{\sqrt{1+(\tilde H')^2}}\right]
	\;-\; \mathrm{Bo}\,\tilde r\,\tilde H \;=\; 0,
	\qquad
	\mathrm{Bo} \equiv \frac{\rho g R^2}{\gamma},
	\label{eq:shape_nonlinear}
\end{equation}
with boundary conditions
\begin{equation}
	\tilde H'(1) \;=\; -\cot\theta,
	\qquad
	\tilde H(\tilde r)\to 0 \quad \text{as}\ \tilde r\to\infty,
	\label{eq:BCs}
\end{equation}
where $\theta$ is the (macroscopic) contact angle at the wall. The Bond number $\mathrm{Bo}$ measures gravity vs.\ surface tension, and is related to the capillary length $\ell_c=\sqrt{\gamma/(\rho g)}$ via
\begin{equation}
	\mathrm{Bo} \;=\; \left(\frac{R}{\ell_c}\right)^2.
	\label{eq:Bo_lc}
\end{equation}

Far from the wall the slope is small, $|\tilde H'|\ll 1$, and the curvature linearizes to $\kappa \approx -(\tilde H''+\tilde H'/\tilde r)$. Equation~\eqref{eq:shape_nonlinear} then reduces to the modified Bessel equation
\begin{equation}
	\tilde H'' + \frac{1}{\tilde r}\,\tilde H' - \mathrm{Bo}\,\tilde H = 0,
	\label{eq:shape_linear}
\end{equation}
whose decaying solution is
\begin{equation}
	\tilde H(\tilde r) \;\propto\; K_0\!\big(\lambda\,\tilde r\big),
	\qquad
	\lambda \equiv \sqrt{\mathrm{Bo}} \;=\; \frac{R}{\ell_c}.
	\label{eq:Bessel_tail}
\end{equation}
Asymptotically, $K_0(\lambda\tilde r)\sim \sqrt{\pi/(2\lambda\tilde r)}\,e^{-\lambda \tilde r}$, i.e.\ an exponential decay with a $1/\sqrt{\tilde r}$ prefactor. Thus, when lengths are scaled by $R$, the inverse decay length measured from the meniscus tail is precisely $\lambda=\sqrt{\mathrm{Bo}}$.

To impose the wall boundary condition $\tilde H'(1)=-\cot\theta$ while retaining the linear Bessel form in the tail, we use
\begin{equation}
	\tilde H(\tilde r) \;=\; C \;+\; \frac{\cot\theta}{\lambda\,K_1(\lambda)}\,K_0\!\big(\lambda\,\tilde r\big),
	\label{eq:model}
\end{equation}
where $C$ is a baseline offset that absorbs small imaging biases in the far field. Using $dK_0(x)/dx=-K_1(x)$, \eqref{eq:model} yields
\begin{equation}
	\tilde H'(1)
	= \frac{\cot\theta}{\lambda K_1(\lambda)}\bigl(-\lambda K_1(\lambda)\bigr)
	= -\cot\theta,
\end{equation}
as required. The fit therefore adjusts only $\lambda>0$ and $C$; the amplitude is fixed by the angle constraint. As soon as $\lambda$ is known, the fitted Bond number follows as
\begin{equation}
	\ \mathrm{Bo}_{\text{fit}}=\lambda^2\ ,
	\qquad
	\ \ell_{c,\text{fit}}=\frac{R}{\lambda}\ .
	\label{eq:Bo_fit_lc_fit}
\end{equation}

\subsection*{Laplace pressure from the fitted shape}
The fitted profile can be used to compute the Laplace pressure jump $\Delta P$ along the interface. Two complementary approaches are used.

\emph{Linearized pressure.} In the linear regime, $\kappa_{\text{lin}}\approx -(\tilde H''+\tilde H'/\tilde r)$ and, using~\eqref{eq:shape_linear}, one obtains
\begin{equation}
	\Delta\tilde P_{\text{lin}}(\tilde r) \;\equiv\; \kappa_{\text{lin}}(\tilde r) \;\approx\; -\,\mathrm{Bo}_{\text{fit}}\,\bigl(\tilde H(\tilde r)-C\bigr).
	\label{eq:Plin}
\end{equation}
The dimensional pressure follows from $\Delta P_{\text{lin}} = (\gamma/R)\,\Delta\tilde P_{\text{lin}}$.

\emph{Exact pressure (nonlinear curvature).} Using the analytic derivatives of~\eqref{eq:model},
\begin{equation}
	\tilde H'(\tilde r) \;=\; -\,\frac{\cot\theta}{K_1(\lambda)}\,K_1\!\bigl(\lambda\tilde r\bigr),
	\qquad
	\tilde H''(\tilde r) \;=\; \frac{\cot\theta\,\lambda}{2K_1(\lambda)}\Bigl[\,K_0\!\bigl(\lambda\tilde r\bigr)+K_2\!\bigl(\lambda\tilde r\bigr)\Bigr],
\end{equation}
substitution into~\eqref{eq:kappa_exact} yields $\kappa_{\text{exact}}(\tilde r)$, and the dimensional pressure jump is
\begin{equation}
	\Delta P_{\text{exact}}(\tilde r) \;=\; \frac{\gamma}{R}\,\kappa_{\text{exact}}(\tilde r).
	\end{equation}}

\end{document}